\documentclass[11pt,onecolumn,amssymb,nofootinbib]{revtex4}
\usepackage{amsmath, amsthm, amscd, amssymb}
\usepackage{bm}
\usepackage{bbm}

\begin{document}

\title{\bf Collapse models with non-white noises II: particle-density coupled noises}
\author{Stephen L. Adler}
\email{adler@ias.edu} \affiliation{Institute for Advanced Study,
Einstein Drive, Princeton, NJ 08540, USA.}
\author{Angelo Bassi}
\email{bassi@ts.infn.it, bassi@mathematik.uni-muenchen.de}
\address{Dipartimento di Fisica Teorica,
Universit\`a di Trieste, Strada Costiera 11, 34014 Trieste, Italy.
\\  Mathematisches Institut der L.M.U., Theresienstr. 39, 80333
M\"unchen, Germany. \\ Istituto Nazionale di Fisica Nucleare,
Sezione di Trieste, Strada Costiera 11, 34014 Trieste, Italy.}
\begin{abstract}
We continue the analysis of  models of spontaneous wave function
collapse with stochastic dynamics driven by non-white Gaussian
noise.  We specialize to a model in which a classical ``noise''
field, with specified autocorrelator, is coupled to a local
nonrelativistic particle density. We derive general results in
this model for the rates of density matrix diagonalization and of
state vector reduction, and show that (in the absence of
decoherence) both processes are governed by essentially the same
rate parameters. As an alternative route to our reduction results,
we also derive the Fokker-Planck equations that correspond to the
initial stochastic Schr\"odinger equation. For specific models of
the noise autocorrelator, including ones motivated by the
structure of thermal Green's functions,  we discuss the qualitative and
qantitative dependence  on
model parameters, with particular emphasis on possible
cosmological sources of the noise field.

\end{abstract}
\maketitle

\section{Introduction}
\label{sec:one}

In an earlier paper~\cite{ab}, hereafter referred to as (I), we
presented a detailed analysis of stochastic models for state
vector collapse driven by Gaussian non-white noise.   In
particular, we showed that a perturbation expansion in the noise
strength parameter $\sqrt{\gamma}$ permits the explicit
calculation of consequences of the model, in parallel with
standard results obtained by the It\^o calculus in the white noise
case.  In (I) the noise couplings were introduced in  generic
form, subject to the assumption that the noise correlator has a
positive definite structure in the large time limit.   As we shall
see, this positivity assumption is overly restrictive, and does
not apply to the physically interesting case of thermal noise,
where the spatial Fourier transform of the noise correlator is
oscillatory in time. Our aim in this paper is to specialize the
discussion of (I) to the physically interesting case of a particle
density-coupled classical noise field, and then within the context
of this model, to give a generalized  analysis of density matrix
diagonalization, state vector reduction, and constraints on model
parameters. We then turn to the question of whether the noise
field postulated in stochastic reduction models can be realized as
a cosmological field.  (For a section-by-section brief summary of
the contents of this paper, the reader should turn to the Summary
and Discussion given in Sec. VII.)

Our starting point in (I) was a diffusion process for the wave
function in Hilbert space having the form (with $\hbar=1$, with
the constant complex coupling factor $\xi$ introduced in (I) set
equal to 1, and with the state vector denoted here by
$|\psi\rangle$),
\begin{equation} \label{eq:cm1}
\frac{d|\psi(t)\rangle}{dt} \quad = \quad \left[ -i H\; +\;
\sqrt{\gamma}\,  \sum_{i=1}^{N} A_{i}w_{i}(t) \; + \; O \right]
|\psi(t)\rangle~~~.
\end{equation}
Here $H$ is the standard quantum Hamiltonian of the system,
$A_{i}$ are commuting self-adjoint operators, $\gamma$ is a
positive coupling constant, and $O$ is a linear operator yet to be
defined.  The noises $w_{i}(t)$ are real Gaussian random
processes, whose mean and correlation functions are, respectively
\begin{equation} \label{eq:cm2}
{\mathbb E} [ w_{i}(t) ] = 0, \qquad {\mathbb E} [ w_{i}(t_{1})
w_{j}(t_{2}) ] = D_{ij}(t_{1},t_{2})~~~.
\end{equation}

We will now specialize the discussion to the case in which the index
$i$ is the spatial coordinate $\vec x$, and the operator $A_i$ is a
particle density $M(\vec x)$, which, for a many body system composed
of distinguishable particles with couplings $m_i$ and coordinate
operators $\vec{q}_i$, is given by
\begin{equation} \label{eq:massden}
M(\vec x)=\sum_i m_i \delta^3(\vec x-\vec q_i)~~~.
\end{equation}
(We have chosen a notation appropriate to the case in which the
density $M$ is a mass density, but \eqref{eq:massden} also
describes other forms of coupling to particle densities, such as
to the baryon number, lepton number,  or isospin densities, with
$m_i$ the appropriate coupling constants.) An important property
of the density operator of \eqref{eq:massden} is that when
integrated over space it reduces to a $c$-number that commutes
with all operators,
\begin{equation} \label{eq:massint}
\int d^3 x M(\vec x)= \int d^3 x \sum_i m_i \delta^3(\vec x-\vec
q_i) = \sum_i m_i .
\end{equation}
Hence a noise coupling to the density operator can be permitted to
have a nonzero expectation, since this will only contribute a
constant term to the effective Hamiltonian on the right of
\eqref{eq:cm1}.   So we will assume that, corresponding to
\eqref{eq:massden}, the noises $w_i(t)$ of (I) form a classical
noise field, which we shall denote by $\phi(\vec x,t)$, with mean
and autocorrelation
\begin{equation} \label{eq:cm3}
{\mathbb E} [ \phi(\vec x,t) ] = \phi_0, \qquad {\mathbb E} [
\big(\phi(\vec x,t_1)-\phi_0)\big) \big(\phi(\vec y,
t_2)-\phi_0\big) ] = D(\vec x-\vec y,t_1-t_2)~~~.
\end{equation}
Here,  in assuming a constant expectation $\phi_0$ and in writing
the arguments of $D$, we have built in an assumption of space and
time translation invariance; we shall also assume spatial
inversion invariance, so that $D(\vec x,t)=D(-\vec x,t)$.   Thus,
with this specialization of the noise structure of (I), the
diffusion process in Hilbert space of \eqref{eq:cm1} becomes
\begin{equation} \label{eq:cm4}
\frac{d|\psi(t)\rangle}{dt} \quad = \quad \left[ -i H\; +\;
\sqrt{\gamma}\,  \int d^3 x M(\vec x) \phi(\vec x,t) \; + \; O
\right] |\psi(t)\rangle~~~.
\end{equation}
In most of what follows, we will neglect the Hamiltonian term in
\eqref{eq:cm4}, focusing on effects that arise from the action of
the stochastic term.

Because it uses real-valued noise,  \eqref{eq:cm4} does not
preserve the norm of the wave function, and this is where the
operator $O$ enters.  In (I), through detailed calculations that
we shall not repeat, we show that $O$ is fixed by the requirements
of (i) state vector normalization, and (ii) a linear evolution
equation for the density matrix
\begin{equation}\label{rhodef}
\rho(t)={\mathbb E}[|\psi(t)\rangle \langle \psi(t)|]~~~,
\end{equation}
which guarantees that superluminal signaling cannot occur.
Relation~\eqref{rhodef} guarantees also the positivity of $\rho(t)$
throughout time.  Determining the structure of $O$ leads to three
equations from (I),  which are exact to order $\gamma$, and
which when specialized to the case of a density coupled noise, form
the starting point for our analysis here.

The first of the needed equations describes the density matrix
time evolution, as given by (53) of (I),
\begin{align}\label{eq:lindnw0}
 \frac{d}{dt} \rho(t) = -i [H,\rho(t)]+& \gamma  \int_0^t
 ds
\int d^3 x \int d^3 y [ M(\vec x)\,\rho(t)\, M(\vec y,s-t) +
M(\vec y,s-t)\,\rho(t) \, M(\vec x)\cr
 -&M(\vec x)M(\vec y,s-t) \rho(t) - \rho(t) M(\vec y, s-t) M(\vec x) ] D(\vec
x-\vec y,t-s)~~~,\cr
\end{align}
with [see (47) of (I)]
\begin{equation}\label{eq:mstdef}
M(\vec y,s-t)=e^{iH(s-t)}M(\vec y\,)e^{-iH(s-t)}~~~.
\end{equation}
When $H=0$ this simplifies to read [see (19) of (I)]
\begin{align}\label{eq:lindnw}
 \frac{d}{dt} \rho(t) = & \gamma \int d^3x
 \int d^3 y [ M(\vec x)\,\rho(t)\, M(\vec y\,) + M(\vec y\,)\,\rho(t)
\, M(\vec x)\cr
 -&M(\vec x)M(\vec y\,) \rho(t) - \rho(t) M(\vec y\,) M(\vec x) ] F(\vec
x-\vec y,t)~~~,\cr
\end{align}
where we have defined
\begin{equation}\label{eq:fdeff}
F(\vec x-\vec y,t) \; = \; \int_0^{t} ds \, D(\vec x-\vec y, t-s).
\end{equation}
To state the second equation, let us define the expectation
$\langle O \rangle_t=\langle \psi(t)|O|\psi(t) \rangle$ for any
operator $O$.  Then when $H=0$ the time evolution of the
stochastic expectation of the variance $V_{A}(t) = \langle A^2
\rangle_{t} - \langle A \rangle_{t}^2$ of any operator $A$ that
commutes with the mass density for all $\vec x$,  given by
(23) and (24) of (I), becomes
\begin{equation} \label{eq:redu}
\frac{d} {dt}{\mathbb E} [ V_{A}(t) ] \quad =  - \; 8\,  \gamma
\,\int d^3 x \int d^3 y  \, {\mathbb E}[ \langle (M(\vec x) -
\langle M(\vec x)\rangle_{t}) A \rangle_{t} \langle (M(\vec y) -
\langle M(\vec y) \rangle_{t}) A \rangle_{t}] \, F(\vec x-\vec
y,t).
\end{equation}
The final equation that we need describes the time evolution of
the state vector $|\psi(t)\rangle$, as specified in (40),
(51), and (52) of (I), which combined become
\begin{equation} \label{eq:cm5}
\frac{d|\psi(t)\rangle}{dt} \quad = \quad \left[ -i H\; +\;
\sqrt{\gamma}\,  \int d^3 x  [M(\vec x) -\langle M(\vec x)
\rangle_t] \phi(\vec x,t) \; + \;\gamma ( B-\langle B
\rangle_t)\right] |\psi(t)\rangle~~~,
\end{equation}
with the self-adjoint operator $B$ given by
\begin{equation} \label{eq:Bdef}
B= -2\int d^3 x \int d^3 y F(\vec x-\vec y,t) [M(\vec x)-\langle
M(\vec x) \rangle_t][M(\vec y)-\langle M(\vec y) \rangle_t]~~~.
\end{equation}
The alternative form of this equation given in (35) and (37)
of (I) differs only by a change of measure for the noise, and
makes identical physical predictions.

As is easily checked, an important consequence of the fact that
the spatial integral of $M(\vec x)$ is a $c$-number (c.f.
\eqref{eq:massint})  is that the noise field expectation $\phi_0$
makes no contribution to the order $\sqrt{\gamma}$ term in
\eqref{eq:cm5}, and that a space-independent constant in $F(\vec
x-\vec y,t)$ makes no contribution to  \eqref{eq:redu},
\eqref{eq:lindnw}, and \eqref{eq:Bdef}.  That is,  we can replace
$F(\vec x-\vec y,t)$ by the subtracted function
\begin{equation}\label{eq:tildef}
 F(\vec x -\vec y,t) - \xi(t)~~~,
 \end{equation}
 for an arbitrary function $\xi(t)$, with no effect on the
 equations; only the nonzero spatial Fourier components of
$F(\vec x-\vec y,t)$ are significant for our analysis.  In
particular, for $\xi(t) =F(\vec 0, t)$, this invariance implies
that we are free to replace $F(\vec x-\vec y,t)$  by the
subtracted function $F(\vec x-\vec y,t)-F(\vec 0,t)$, which has a
spatial Fourier transform with improved convergence at small wave
numbers.

There has recently been a spirited debate ~\cite{rel} over whether
stochastic reduction models can be made relativistically
invariant.  We remark in this context that the noise coupling of
\eqref{eq:cm1} can be obtained in a number of ways as the
non-relativistic limit of relativistically invariant,
anti-self-adjoint coupling actions involving scalar, vector, or
tensor fields.  (An  anti-self-adjoint action is required to give
a real noise term in the Schr\"odinger equation; we will not
attempt here a fundamental justification of this
phenomenologically-motivated choice of Hermiticity structure.)
When the noise coupling is introduced as the nonrelativistic limit
of a relativistic action, relativistic invariance of the
stochastic reduction model is broken not by the noise coupling,
but by the assumed autocorrelator of the noise field $\phi(\vec
x,t)$.  For example, if the noise field has a cosmological origin,
 its autocorrelator might  be expected to refer preferentially to
either the Lorentz frame in which the cosmological background
radiation is isotropic, or to the galactic rest frame.    A topic
for future work will be to investigate whether an effective
anti-self-adjoint coupling action can arise naturally in a
non-equilibrium cosmology, or requires an explicitly non-unitary
pre-quantum dynamics.

\section{Density matrix diagonalization}
\label{sec:two}

We begin our analysis by considering the consequences of
\eqref{eq:lindnw} for coordinate off-diagonal matrix elements of
the density matrix, when the Hamiltonian evolution is neglected.
Taking the matrix element of \eqref{eq:lindnw} between states
$\langle \{\vec r_{\ell}^1\}|$ and $|\{\vec r_{\ell}^2\} \rangle$,
we get a differential equation for the time dependence of the
matrix element of $\rho$, which can be immediately integrated to
give
\begin{equation}\label{eq:densdecay}
\langle \{\vec r_{\ell}^1\}|\rho(t)|\{\vec r_{\ell}^2\} \rangle
=e^{-\Gamma(t)}\langle \{\vec r_{\ell}^{\,1}\}|\rho(0)|\{\vec
r_{\ell}^{\,2}\} \rangle~~~,
\end{equation}
with the integrated rate $\Gamma(t)$ given by
\begin{equation}\label{eq:gamdef}
\Gamma(t)=\gamma \int d^3x \int d^3 y \int_0^t ds F(\vec x-\vec
y,s) [m_1(\vec x)-m_2(\vec x)] [m_1(\vec y)-m_2(\vec y)]~~~,
\end{equation}
where $m_{1,2}$ are the eigenvalues of the operator $M(\vec x)$
when acting on the respective states $|\{\vec
r_{\ell}^{\,1,2}\}\rangle$,
\begin{equation}\label{eq:m12def}
m_1(\vec x)=\sum_i m_i \delta^3(\vec x -\vec r_i^{\,1})~,~~
m_2(\vec x)=\sum_i m_i \delta^3(\vec x -\vec r_i^{\,2})~~~.
\end{equation}
Substituting \eqref{eq:m12def} into \eqref{eq:gamdef} and carrying
out the $\vec x$ and $\vec y$ integrals using the delta functions,
we obtain
\begin{equation}\label{eq:gamnew}
\Gamma(t)=\gamma \int_0^t ds  \sum_i\sum_j m_i m_j [F(\vec
r_i^{\,1}-\vec r_j^{\,1},s) +F(\vec r_i^{\,2}-\vec
r_j^{\,2},s)-F(\vec r_i^{\,1}-\vec r_j^{\,2},s) - F(\vec
r_i^{\,2}-\vec r_j^{\,1},s)]~~~.
\end{equation}

We now review a number of useful features of this formula (many of
which, in a slightly different notation, are familiar from the
stochastic reduction literature).   First of all, as already
pointed out in Sec. 1, \eqref{eq:gamnew} is unchanged when we
replace $F(\vec r,s)$ by the subtracted function $F(\vec
r,s)-F(\vec 0,s)$. Secondly, suppose that $\vec r_I^{\,1}=\vec
r_I^{\,2}=\vec r_I$ for some particle with index $I$.  Then the
contribution of this particle to the double sum in
\eqref{eq:gamnew} is
\begin{equation}
2m_I^2[F(\vec 0,s)-F(\vec 0,s)] + 2m_I \sum_{j \neq I} m_j [F(\vec
r_I-\vec r_j^{\,1},s) +F(\vec r_I-\vec r_j^{\,2},s)-F(\vec
r_I-\vec r_j^{\,2},s) - F(\vec r_I-\vec r_j^{\,1},s)]=0~~~.
\end{equation}
So only particles that have different coordinates in the groups 1
and 2 contribute to the sum.

Third, suppose that for large separations $\vec r$, relative to
some  correlation scale $r_C$, the function $F(\vec r,s)$
asymptotically approaches a constant (which can be zero or
nonzero).  Then if there are two particles $I,J$ such that $\vec
r_I^{\,1,2}-\vec r_J^{\,1,2}$ are all large enough relative to
$r_C$ to be in the asymptotic regime for $F$,  the cross terms in
the double sum linking these two particles do not contribute. This
means that if the particles form a set of $K$ widely spaced
bunches on the scale of $r_C$, with the particles of group 2
displaced with respect to those of group 1 by distances of order
$r_C$, the formula for $\Gamma(t)$ splits into a sum
\begin{equation}\label{eq:split}
\Gamma(t)=\sum_{k=1}^K \Gamma^k(t)~~~,
\end{equation}
with $\Gamma^k(t)$ computed entirely within the $k$th bunch.

Fourth, let us take group 1 to be a collection of particles that
are very closely spaced on the scale of $r_C$, and suppose that
the particles of group 2 are all displaced by a common vector
$\vec R$ with respect to those of group 1.  In this case,
$\Gamma(t)$ is approximated by the formula
\begin{equation} \label{eq:cm}
\Gamma(t)\simeq 2 \gamma \int_0^t ds \left( \sum_i m_i \right)^2
[F(\vec 0,s) - F(\vec R,s)]~~~,
\end{equation}
which is the formula that would be obtained if there were only one
particle of mass $\sum_i m_i$ at the center of mass of the group.
 The above formulas display the amplification mechanism typical
of collapse models: when particles interact to form a macro-object,
the collapses on the single particles add up in such a way that the
center of mass of the object collapses each time a single particle
does. This is the reason why these models can account both for the
quantum properties of microscopic systems and for the classical
properties of macroscopic objects.

Fifth,  let us again take group 1 to be a collection of particles
that are very closely spaced on the scale of $r_C$, but now suppose
that the particles of group 2 are displaced by random amounts, with
an average magnitude of displacement $R$ with respect to those of
group 1.  When the function $F(\vec r,s)$ only depends on the
magnitude $|\vec r|$ of the displacement vector, so that $F(\vec
r,s)=F[|\vec r|,s]$ then $\Gamma(t)$ is approximated by the formula
\begin{equation} \label{eq:rand1}
\Gamma(t) \simeq \gamma \int_0^t ds \left( \sum_i m_i \right)^2
\big[F(0,s) + \langle\!\langle F(|\vec r_i^{\,2}-\vec r_j^{\,2}|,s)
\rangle\!\rangle_{N}- 2\langle\!\langle F(|\vec r_i^{\,1}-\vec
r_j^{\,2}|,s)\rangle\!\rangle_{N}\big]~~~,
\end{equation}
 where $\langle\!\langle ... \rangle\!\rangle_{N}$ denotes the
average over the ensemble of particles; when $R>r_C$
\eqref{eq:rand1}  is further approximated by
\begin{equation}\label{eq:rand2}
\Gamma(t)\simeq \gamma \int_0^t ds \left( \sum_i m_i \right)^2
\big[F(0,s) - F(R,s)\big]~~~,
\end{equation}
which is one half of the $\Gamma(t)$ given by the center of mass
formula \eqref{eq:cm} for the corresponding magnitude of $R$.

Finally, in many cases of interest $F(\vec x-\vec y,s)$ can be
written as a sum or integral over factors referring to $\vec x$
and $\vec y$ separately,
\begin{equation}\label{eq:factor}
F(\vec x-\vec y,s)=\sum_{\alpha} {\cal F}(\alpha,s) f(\alpha,\vec
x) f(\alpha,\vec y)~~~,
\end{equation}
with ${\cal F}$ an appropriate weighting function, and $\alpha$ a
shorthand for any combination of discrete and continuous
variables.  Substitution of \eqref{eq:factor} into
\eqref{eq:gamdef} gives
\begin{equation}\label{gamfactored}
\Gamma(t)=\gamma \int_0^t ds \sum_{\alpha}{\cal F}(\alpha,s)
g(\alpha,\{\vec r_{\ell}^{\,1}\},\{\vec r_{\ell}^{\,2}\})^2~~~,
\end{equation}
with
\begin{equation}\label{eq:gdef}
g(\alpha,\{\vec r_{\ell}^{\,1}\},\{\vec r_{\ell}^{\,2}\})=\int
d^3x f(\alpha,\vec x) [m_1(\vec x)-m_2(\vec x)] = \sum_i m_i
[f(\alpha,\vec r_i^{\,1}) - f(\alpha,\vec r_i^{\,2})]~~~.
\end{equation}

\section{State vector reduction}
\label{sec:three}

We proceed next to apply \eqref{eq:redu} to the problem of
reduction of a state vector constructed as a superposition of
position eigenstates $|\{\vec
r_{\ell}^{\,j}\}\rangle~,~~j=1,...,N$. Our first step is to
rewrite \eqref{eq:redu} in a more useful form by setting $A=B+C$,
with $B$ and $C$ operators that commute with each other and with
the mass density, and subtracting off \eqref{eq:redu} as written
for $B$ and $C$ alone, which gives
\begin{equation} \label{eq:redubc}
\frac{d} {dt}{\mathbb E} [\langle BC \rangle_t - \langle B
\rangle_t \langle C \rangle_t ] = - \; 8\, \gamma \,\int d^3 x
\int d^3 y \, {\mathbb E}[ \langle (M(\vec x) - \langle M(\vec
x)\rangle_{t}) B \rangle_{t} \langle (M(\vec y) - \langle M(\vec
y) \rangle_{t}) C \rangle_{t}] \, F(\vec x-\vec y,t)~~~.
\end{equation}
Using the fact that $B$ and $C$ can be arbitrary operator functions
of the particle coordinate operators, by making the specific choices 
 $B = \prod_{i} \delta^3(\vec u_i-\vec q_i)$ and $C =
\prod_{i} \delta^3(\vec w_i-\vec q_i)$, we get  
\begin{align}\label{eq:sqevol}
~&\frac{d}{dt}{\mathbb E} \Bigg[\Bigg(\prod_i \delta^3(\vec
w_i-\vec u_i)\Bigg) |\psi(\{\vec w_{\ell}\})|^2 - |\psi(\{\vec
u_{\ell}\})|^2 |\psi(\{\vec w_{\ell}\})|^2 \Bigg] \cr
 =&-8\gamma\int d^3x d^3y {\mathbb
E}\Bigg[|\psi(\{\vec u_{\ell}\})|^2 | \psi(\{\vec w_{\ell}\})|^2\cr
\times & \sum_j m_j[\delta^3(\vec x-\vec u_j)-|\hat \psi_j(\vec
x)|^2] \sum_k m_k [\delta^3(\vec y-\vec w_k)-|\hat \psi_k(\vec
y)|^2]\Bigg] F(\vec x -\vec y,t) ~~~,\cr
\end{align}
where we have introduced the definition
\begin{align}\label{eq:hatpsidef}
|\hat \psi_j(\vec z_j)|^2=&\Bigg(\prod_{ i \neq j}\int d^3
z_i\Bigg) |\psi(\{\vec z_{\ell}\})|^2~~~,\cr \int d^3 z_j |\hat
\psi(\vec z_j)|^2=&1~~~.\cr
\end{align}

Let us now specialize \eqref{eq:sqevol} to the case of a wave
function which is the superposition of $N$ distinct localized
groups of particles, by writing
\begin{align}\label{eq:psisup}
\psi(\{\vec z_{\ell}\})= \langle \{\vec z_{\ell} \}|\psi(t)\rangle
= &\sum_{J=1}^N \alpha_J \prod_{\ell} \delta^3(\vec
z_{\ell}-\vec r_{\ell}^J)^{1/2}~~~,\cr |\psi(\{\vec
z_{\ell}\})|^2=&\sum_{J=1}^N p_J \prod_{\ell} \delta^3(\vec
z_{\ell}-\vec r_{\ell}^J)~~~,\cr
\end{align}
with $p_J=|\alpha_J|^2 $ and with normalization of the wave function
implying that $\sum_J p_J=1$. (By the square root of a delta
function, we mean a Gaussian wave packet which is sharply localized,
with a modulus squared that integrates to unity.) Substituting
\eqref{eq:psisup} into \eqref{eq:sqevol},  and integrating in $d
\{ \vec w_{\ell} \}$ around $\{ \vec r_{\ell}^{\,L} \}$  and in $d
\{ \vec u_{\ell} \}$ around $\{ \vec r_{\ell}^{\,M} \}$,  we get an
equation for the time evolution of the occupation probabilities $p_J$ of
the corresponding states $\prod_{\ell}\delta^3(\vec z_{\ell}-\vec r_{\ell}^J)^{1/2}$ with label $J$  that appear in the superposition,
\begin{align}\label{eq:probevol}
\frac{d}{dt}{\mathbb E}[\delta_{ML}p_M-p_Mp_L]=&-8\gamma {\mathbb
E}\Bigg[p_Mp_L \sum_j \sum_k m_j m_k \Bigg\{ F(\vec r_j^{\,L}-\vec
r_k^{\,M},t) + \sum_{R,\,S}p_R p_S F(\vec r_j^{\,R}-\vec
r_k^{\,S},t) \cr -& \sum_R p_R F(\vec r_j^{\,R}-\vec
r_k^M,t)-\sum_{\,S} p_S F(\vec r_j^{\,L}-\vec r_k^{\,S},t) \Bigg\}
\Bigg]~~~.
\end{align}
Specializing this further to the two group case with $N=2$, taking
$M=L=1$ and doing some algebraic rearrangement using the fact that
the sum of the probabilities is $p_1+p_{\,2}=1$, we get
\begin{align}\label{eq:twoevol}
\frac{d}{dt}{\mathbb E}[p_1p_{\,2}]=&-8\gamma {\mathbb
E}[p_1^2p_{\,2}^2] \sum_j\sum_k m_j m_k \{ F(\vec r_j^{\,1}-\vec
r_k^{\,1},t) + F(\vec r_j^{\,2}-\vec r_k^{\,2},t)\cr
 -&F(\vec
r_j^{\,1}-\vec r_k^{\,2},t) -F(\vec r_j^{\,2}-\vec r_k^{\,1},t) \}
~~~.\cr
\end{align}

We can now use \eqref{eq:twoevol} to derive upper and lower bounds
to the reduction rate, as follows.  To obtain an upper bound, we
use the inequality
\begin{equation}\label{eq:jensen}
{\mathbb E}[p_1^2 p_{\,2}^2] \geq {\mathbb E}[p_1p_{\,2}]^2~~~,
\end{equation}
and the assumption that the integrand of $\Gamma(t)$ in
\eqref{eq:gamnew} is  positive for all $s$, to rewrite
\eqref{eq:twoevol} as
\begin{align}\label{eq:twoevol1}
\frac{d}{dt}{\mathbb E}[p_1p_{\,2}]\leq&-8\gamma {\mathbb
E}[p_1p_{\,2}]^2 \sum_j\sum_k m_j m_k \{ F(\vec r_j^{\,1}-\vec
r_k^{\,1},t) + F(\vec r_j^{\,2}-\vec r_k^{\,2},t)\cr
 -&F(\vec
r_j^{\,1}-\vec r_k^{\,2},t) -F(\vec r_j^{\,2}-\vec r_k^{\,1},t) \}
~~~,\cr
\end{align}
giving a differential inequality that can be integrated to give an
upper bound on the reduction rate
\begin{equation}\label{eq:upperbound}
{\mathbb E}[p_1(t)p_{\,2}(t)] \leq \frac { {\mathbb
E}[p_1(0)p_{\,2}(0)] } {1 + 8 \Gamma(t)}~~~.
\end{equation}
To get a lower bound, we note that since the probabilities $p_1$
and $p_2$ obey $p_1+p_2=1$, we have $p_1 p_2=p_1(1-p_1) \leq 1/4$,
and so
\begin{equation}\label{eq:probbound}
{\mathbb E}[p_1^2 p_{\,2}^2] \leq {\mathbb E}[p_1 p_{\,2}]/4~~~.
\end{equation}
Again assuming that the integrand of \eqref{eq:gamnew} is
positive for all $s$, this gives a differential inequality that can be
integrated to give the lower bound
\begin{equation}\label{eq:lower1}
{\mathbb E}[p_1(t)p_{\,2}(t)] \geq {\mathbb E}[p_1(0)p_{\,2}(0)]
\exp[-2\Gamma(t)]~~~.
\end{equation}
Thus we see that in our model of a Schr\"odinger equation modified
solely  by a real noise process, the upper and lower bounds on the
reduction rate involve (under the uniform positivity assumption)
the same integrated rate function $\Gamma(t)$ as appears in the
decay of the off-diagonal density matrix element $\langle \{
\vec r_{\ell}^{\,1}\}|\rho(t)|\{ \vec r_{\ell}^{\,2}\} \rangle$.  Of course, in
realistic applications, the rate for density matrix
diagonalization is expected to receive much larger contributions
from  decoherence processes, which can be modeled by imaginary
noise terms in the Schr\"odinger equation that do not contribute
to state vector reduction.  Although the upper and lower bounds
are governed by the same integrated rate function, they have very
different functional dependencies: the upper bound depends on the
inverse of $\Gamma(t)$, whereas the lower bound is a negative
exponential in $\Gamma(t)$.  Solvable models \cite{bass}, \cite
{hugh}, and Appendix D, show that in fact the actual decay of the variance is
exponential, rather than power law, indicating that the lower
bound of \eqref{eq:lower1} gives the better estimate.\footnote{
In the example calculated in Appendix D, the actual variance decay is
$\sim e^{-\Gamma(t)}$.  A
simple example shows how an exponential decay of the variance can
agree with the inequalities used to get the upper and lower
bounds.  If ${\mathbb E}[p_1(t) p_2(t)] \simeq  {\mathbb
E}[p_1(0)p_2(0)] \exp(-\Gamma(t))$, then $(d/dt){\mathbb E}
[p_1(t)p_2(t)]=- \Gamma^{\prime}(t){\mathbb E} [p_1(t)p_2(t)]$,
whereas \eqref{eq:twoevol} implies that $(d/dt){\mathbb E}
[p_1(t)p_2(t)]=-8 \Gamma^{\prime}(t){\mathbb E}
[p_1^2(t)p_2^2(t)]$, and so we must have ${\mathbb E}[p_1(t)
p_(t)]=8{\mathbb E}[ p_1^2 p_2^2]$.  Suppose now that
$p_1(t)p_2(t)=0$ with probability $1-\exp(-\Gamma(t))$, and
$p_1(t)p_2(t)=1/8$ with probability $\exp(-\Gamma(t))$. We then
have ${\mathbb E}[p_1(t)p_2(t)]=8{\mathbb
E}[p_1^2(t)p_2^2(t)]=(1/8)\exp(-\Gamma(t))$.  However, ${\mathbb
E}[p_1(t)p_2(t)]^2 =(1/64) \exp(-2\Gamma(t))$, which for large
$\Gamma(t)$ is much smaller than ${\mathbb
E}[p_1^2(t)p_2^2(t)]=(1/64) \exp(- \Gamma(t))$, and so the
inequality of \eqref{eq:jensen}, which was  used to get the upper
bound, is far from being saturated, while by construction, the
inequality of \eqref{eq:probbound}, which was used to get the
lower bound, is saturated. }

In the general $N$ group case, although we have not derived
rigorous bounds, we can get estimates similar to the two-group
case by setting $M=L$ in \eqref{eq:probevol}, giving
\begin{align}\label{eq:probevol1} \frac{d}{dt}{\mathbb
E}[p_L(1-p_L)]=&-8\gamma {\mathbb E}\Bigg[p_L^2 \sum_j \sum_k m_j
m_k  \Bigg\{ F(\vec r_j^{\,L}-\vec r_k^{\,L},t) + \sum_{R,\,S}p_R
p_S F(\vec r_j^{\,R}-\vec r_k^{\,S},t) \cr -& \sum_R p_R F(\vec
r_j^{\,R}-\vec r_k^{\,L},t)-\sum_{\,S} p_S F(\vec r_j^{\,L}-\vec
r_k^{\,S},t) \Bigg\} \Bigg]~~~.
\end{align}
Suppose now that the stochastic process brings the probabilities
close to  a corner of their domain, where for some $M\not= L$ the
probability $p_M$ is close to unity, and thus all the other
probabilities are small.  The right-hand side of
\eqref{eq:probevol1} then contains terms of second degree in small
quantities, given by selecting the terms with $R=M$ and $S=M$ in
the sums, plus remaining terms that are third degree in small
quantities.  The second degree terms contribute
\begin{align}\label{eq:probevol2} \frac{d}{dt}{\mathbb
E}[p_L]\simeq &-8\gamma {\mathbb E}[p_L^2] \sum_j \sum_k m_j m_k
\Bigg\{ F(\vec r_j^{\,L}-\vec r_k^{\,L},t) + F(\vec r_j^{\,M}-\vec
r_k^{\,M},t) \cr -&  F(\vec r_j^{\,M}-\vec r_k^{\,L},t)-F(\vec
r_j^{\,L}-\vec r_k^{\,M},t) \Bigg\} ~~~,
\end{align}
which has a  structure similar to \eqref{eq:twoevol} for the
two-group case.  Using the inequality
\begin{equation}\label{eq:jensen1}
{\mathbb E}[p_L^2] \geq {\mathbb E}[p_L]^2~~~,
\end{equation}
defining $\Gamma^{LM}(t)$ by
\begin{equation}\label{eq:gamlm}
\Gamma^{LM}(t)=\gamma \int_0^t ds \sum_j\sum_k m_j m_k [F(\vec
r_j^{\,L}-\vec r_k^{\,L},s) +F(\vec r_j^{\,M}-\vec
r_k^{\,M},s)-F(\vec r_j^{\,L}-\vec r_k^{\,M},s) - F(\vec
r_j^{\,M}-\vec r_k^{\,L},s)]~~~,
\end{equation}
and assuming positivity of the integrand of \eqref{eq:gamlm}, we get
a differential inequality that can be integrated to give an upper
bound on the decay rate,
\begin{equation}\label{eq:upperbound1}
{\mathbb E}[p_L(t)] \leq \frac { {\mathbb E}[p_L(0)] } {1 + 8
\Gamma^{LM}(t)}~~~.
\end{equation}
Similarly, from \eqref{eq:probevol2} we can also get a lower bound
on the decay rate,
\begin{equation}\label{lower2}
 {\mathbb E}[p_L(t)] \geq {\mathbb
E}[p_L(0)] \exp[-2\Gamma^{LM}(t)]~~~.
\end{equation}
Thus, near the corner where $p_M\simeq 1$, the other $p_L$ decay
to zero, with the slowest rate of decrease corresponding to the
smallest value of $\Gamma^{LM}$ for $L\not=M$.

To conclude this section, we note that when $F(\vec x-\vec y,t)$
has the factorized form given in \eqref{eq:factor}, then
\eqref{eq:probevol} takes the form
\begin{align}\label{eq:facevol}
\frac{d}{dt}{\mathbb E}[\delta_{ML}p_M-p_Mp_L]=&-8\gamma {\mathbb
E}\Bigg[p_Mp_L \sum_{\alpha} {\cal F} (\alpha,t) \sum_j m_j
[f(\alpha,\vec r_j^{\,L})-\sum_R p_R f(\alpha,\vec r_j^{\,R})]\cr
\times &\sum_k m_k [f(\alpha,\vec r_k^{\,M})-\sum_S p_S
f(\alpha,\vec r_k^{\,S})] \Bigg]~~~,\cr
\end{align}
while \eqref{eq:gamlm} becomes
\begin{align}\label{gamlmfac}
\Gamma^{LM}(t)=&\gamma \int_0^t ds \sum_{\alpha} {\cal F}
(\alpha,t) [\sum_i m_i \big(f(\alpha, \vec r_i^{\,L})-f(\alpha,
\vec r_i^{\, M}) \big)]^2\cr
 =&\gamma \int_0^t ds \sum_{\alpha}
{\cal F}(\alpha,t) g(\alpha,\{\vec r_{\ell}^{\,L}\},\{\vec
r_{\ell}^{\,M}\})^2~~~,\cr
\end{align}
and $\Gamma(t)=\Gamma^{12}(t)$ is the specialization of this
formula to $L=1,~M=2$.

\section{Fokker-Planck equation} \label{sec:four}

As a complement to the methods used in the preceding sections, we
derive the Fokker-Planck equation for the non-white noise model,
and use it to rederive \eqref{eq:probevol}.  We again restrict
ourselves to the case when the Hamiltonian $H$ is zero, which
allows all equations to be diagonalized in coordinate
representation.  Starting from \eqref{eq:cm5} and substituting
\begin{equation}\label{eq:psisup1}
|\psi(t)\rangle = \sum_{L=1}^N \alpha_L
 |\{\vec r_{\ell}^{\,L}\}\rangle~~~,
\end{equation}
with $|\{\vec r_{\ell}^{\,L}\}\rangle$ sharply localized wave-packet
states [c.f. \eqref{eq:psisup}], we find that the coefficients
$\alpha_L$ obey the equation of motion
\begin{equation}\label{eq:alph1}
\frac {d}{dt}\alpha_L = \alpha_L X_L~~~,
\end{equation}
with $X_L$ given by
\begin{align}\label{eq:alph2}
X_L= & \sqrt{\gamma}\sum_i m_i[ \phi(\vec r_i^{\,L},t)-\sum_R p_R
\phi(\vec r_i^{\, R},t) ]  \cr -&\gamma \sum_i \sum_j  2m_i m_j
\Bigg[F(\vec r_i^{\,L}-\vec r_j^{\,L},t)+2\sum_{R,\,S} p_R p_S
F(\vec r_i^{\,R}-\vec r_j^{\,S},t) \cr
 -&2\sum_Rp_R F(\vec
r_i^{\,L}-\vec r_j^{\,R},t)-\sum_Rp_R F(\vec r_i^{\, R}-\vec
r_j^{\, R},t) \Bigg]~~~.\cr
\end{align}
Since $X_L$ is real, and $p_L=\alpha_L^*\alpha_L$, we
correspondingly have
\begin{equation}\label{eq:p1}
\frac {d}{dt}p_L = 2 p_L X_L~~~.
\end{equation}

In order to derive the Fokker-Planck equation, we have to evaluate
${\mathbb E}[(d/dt) f(\{p_R\})]$ for an arbitrary function $f$ of
the set of probabilities $\{p_R\}$, keeping terms through order
$\gamma$.  On using the chain rule we have
\begin{equation}\label{eq:chain}
{\mathbb E}\Bigg[\frac {d}{dt} f(\{p_R\})\Bigg]= {\mathbb
E}\Bigg[\sum_S \frac {\partial f(\{p_R\})}{\partial p_S} \frac
{d}{dt}p_S\Bigg]~~~.
\end{equation}
On substituting \eqref{eq:p1} and \eqref{eq:alph2} for
$(d/dt)p_S$, we encounter two types of terms.  Terms of the form
\begin{equation}\label{eq:type1}
-\gamma \sum_S  {\mathbb E}\Bigg[\frac {\partial
f(\{p_R\})}{\partial p_S} A_S\Bigg]
\end{equation}
can be read off directly from the term proportional to $-\gamma$
in  $X_S$, while terms of type
\begin{equation}\label{eq:type2}
\sqrt{\gamma}\sum_{i,\,S,\,L}{\mathbb E} \Bigg[ \frac {\partial
f(\{p_R\}) }{\partial p_S} B_{S\, i\, L} \phi(\vec r_i^{\,L},t)
\Bigg]
\end{equation}
are evaluated using the Furutsu-Novikov formula, which
approximated by using $p_R(s)=p_R(t) + O(\surd \gamma)$, takes the
form
\begin{equation}\label{eq:fn}
{\mathbb E} \Bigg[ \frac {\partial f(\{p_R\}) }{\partial p_S}
B_{S\, i\, L} \phi(\vec r_i^{\,L},t) \Bigg] =\sum_{j,\,S,\,T}
F(\vec r_i^{\, L}-\vec r_j^{\, S},t) {\mathbb E} \Bigg[
\frac{\partial} {\partial p_T} \Bigg(\frac {\partial f(\{p_R\})
}{\partial p_S} B_{S\, i\, L}\Bigg)  \frac {\partial p_T}
{\partial \phi(\vec r_j^{\, S},t)}\Bigg]~~~.
\end{equation}
The needed derivative of $p_T$ can be read off directly from the
$\sqrt{\gamma}$ term in $X_T$; substituting this, and doing much
algebra, one finds that all first derivatives of $f$ with respect
to the probabilities cancel exactly, leaving finally the compact
expression
\begin{equation}\label{eq:expf} \frac{d}{dt} {\mathbb
E}[ f(\{p_R\})]= 4\gamma \sum_{M,\,T,\,i,\,j,\,Q,\,S}F(\vec
r_i^{\,R}-\vec r_j^{\, S},t) m_i m_j  {\mathbb E} \Bigg[
\frac{\partial^2 f(\{p_R\})}{\partial p_N \partial p_M} p_T p_M
(\delta_{MQ}-p_Q)(\delta_{TS}-p_S)\Bigg]~~~.
\end{equation}

Introducing the probability density $P(\{p_R\},t)$, which includes
as a factor the constraint $\delta(\sum_L p_L-1)$ requiring that
the probabilities sum to unity,  we can also write the expectation
of $(d/dt)f(\{p_R\})$ as
\begin{equation}\label{eq:expect}
\frac{d}{dt} {\mathbb E}[ f(\{p_R\})]=\prod_L \int dp_L
 \frac {\partial P(\{p_R\})}{\partial t} f(\{p_R\})~~~.
 \end{equation}
 Comparison of this expression with \eqref{eq:expf}, as rearranged by two
 integrations by parts  (the surface terms when any probability is
 0 or 1 do not to contribute; see below), one gets the
 Fokker-Planck equation,
\begin{equation}\label{eq:fokpl}
\frac {\partial P(\{p_R\})}{\partial t} = \sum_{M,\,T}
\frac{\partial^2}{\partial p_M \partial p_T} [A_{MT}(\{p_R\})
P(\{p_R\},t)]~~~,
\end{equation}
with
\begin{align}\label{eq:adef}
A_{MT}=& 4\gamma p_M p_T\sum_{i,\,j,\,Q,\,S}F(\vec r_i^{\, R}-\vec
r_j^{\, S},t) m_i m_j (\delta_{MQ}-p_Q)(\delta_{TS}-p_S)\cr =&
4\gamma p_M p_T \sum_{i,\,j,\,Q,\,S}  m_i m_j p_Q p_S [F(\vec
r_i^{\, M}-\vec r_j^{\, T},t)+F(\vec r_i^{\, Q}-\vec r_j^{\, S},t)
-F(\vec r_i^{\, Q}-\vec r_j^{\, T},t) -F(\vec r_i^{\, M}-\vec
r_j^{\, S},t)]~~~. \cr
\end{align}
This equation is a specific case of a general Fokker-Planck
equation written down by Pearle ~\cite{pearle} as the basis for a
general class of objective reduction models.  When $F(\vec x-\vec
y,t)$ takes the factorized form of \eqref{eq:factor}, $A_{MT}$ can
be rewritten as
\begin{equation}\label{afactored}
A_{MT}=4\gamma p_M p_T \sum_{\alpha} {\cal F}(\alpha,t) \phi_M
\phi_T~~~,
\end{equation}
with
\begin{equation}\label{eq:phidef}
\phi_M=\sum_{i,\,Q} m_i p_Q[f(\alpha,\vec r_i^{\, M})-f(\alpha,
\vec r_j^{\, Q})]~~~.
\end{equation}
We see that in addition to vanishing when either $p_M=0$ or
$p_T=0$,  $A_{MT}$ vanishes when either $p_M=1$ or $p_T=1$,
because $\phi_M$ vanishes when $p_M=1$.  This is why the
integrations by parts leading to the Fokker-Planck equation
produce no surface terms, and also why the Fokker-Planck equation
of \eqref{eq:fokpl} satisfies the criteria that Pearle \cite{
pearle} formulated for getting a Fokker-Planck equation that leads
to state vector reduction with Born rule probabilities.

As an application of \eqref{eq:expf}, if we substitute $
f(\{p_R\})=p_L$ we find, since $\partial^2 p_L/(\partial p_N
\partial p_M)=0$, that
\begin{equation}\label{eq:pexp}
\frac{d}{dt} {\mathbb E}[p_L]=0~~~.
\end{equation}
Similarly, if we substitute $f(\{p_R\})=\delta_{KL}p_L-p_Kp_L$, we
find, using
\begin{equation}\label{eq:deriv}
\frac{\partial^2}{\partial p_M \partial
p_N}[\delta_{KL}p_L-p_Kp_L]=-[\delta_{MK}\delta_{NL}
+\delta_{ML}\delta_{NK}]~~~,
\end{equation}
that \eqref{eq:expf} yields \eqref{eq:probevol}.  More generally,
\eqref{eq:expf} and the corresponding Fokker-Planck equation of
\eqref{eq:fokpl} allow one to calculate the time evolution of a
general function $f(\{p_R\})$ of the probabilities.

\section{Noise effects:  Energy production and radiation
by atoms} \label{sec:five}

The noise coupling postulated in Sec. 1 as the origin of state
vector reduction  has other physical effects, that serve to place
upper bounds on the noise coupling strength $\gamma$. We focus in
this section in particular on energy production, and gamma
radiation from atoms, which place particularly stringent bounds on
the model parameters.

\subsection{Energy production}

To calculate the mean rate of energy production, we have to
evaluate $(d/dt){\rm Tr} H \rho(t) ={\rm Tr} H (d/dt) \rho(t)$.
{}From \eqref{eq:lindnw0} we find, by repeated cyclic permutation
under the trace, that
\begin{equation}\label{eq:enprod}
\frac{d}{dt}{\rm Tr}H\rho(t)=-\gamma \int d^3x \int d^3y \int_0^t ds
D(\vec x - \vec y, t-s) {\rm Tr}\big([[H,M(\vec x)],M(\vec
y,s-t)]\rho(t) \big)~~~.
\end{equation}
This equation is exact through order $\gamma$. We now make the
Markovian approximation, of assuming that we can ignore the
``memory effect'' associated with the characteristic decay time of
the noise correlator $D(\vec x-\vec y, t-s)$, by replacing $M(\vec
y,s-t)$ by $M(\vec y,0)=M(\vec y)$. For white noise, where $D(\vec
x-\vec y, t-s) =G(\vec x-\vec y) \delta(t-s)$, the Markovian
approximation is exact; for non-white thermal noises, it should be
a good approximation when the energy at the peak of the noise
spectrum is much higher than the typical kinetic energies  of the
particles to which the noise couples (see Appendix A). With this
approximation, \eqref{eq:enprod} simplifies to
\begin{equation} \label{eq:enprod1} \frac{d}{dt}{\rm
Tr}H\rho(t)=-\gamma \int d^3x \int d^3y F(\vec x - \vec y, t) {\rm
Tr}\big([[H,M(\vec x)],M(\vec y)]\rho(t) \big)~~~,
\end{equation}
where we have made use of the definition \eqref{eq:fdeff}.

Let us now assume that $H$ is the nonrelativistic Hamiltonian for
a collection of particles interacting through a general
velocity-independent potential,
\begin{equation}\label{eq:hsysdef}
H=\sum_i \frac {\vec p_i^{\,2}}{2M_i} + V(\{\vec q_{\ell}\})~~~,
\end{equation}
while $M(\vec x)$ has the form of \eqref{eq:massden} (in the
mass-coupled noise case, $m_i=M_i$) and $F(\vec x-\vec y,t)$ has
the factor decomposition of \eqref{eq:factor}.  Then carrying out
the $\vec x$ and $\vec y$ integrals, \eqref{eq:enprod1} becomes
\begin{equation} \label{eq:enprod2} \frac{d}{dt}{\rm
Tr}H\rho(t)=-\gamma \sum_{\alpha}{\cal F}(\alpha,t)\sum_{i,\,j} m_i
m_j  {\rm Tr}\big([[H,f(\alpha, \vec q_i)],f(\alpha,\vec
q_j)]\rho(t) \big)~~~.
\end{equation}
The commutators appearing in \eqref{eq:enprod2} are easily
evaluated,
\begin{align}\label{eq:comm1}
[H,f(\alpha,\vec q_i)]=&\Bigg[\sum_j\frac {\vec
p_j^{\,2}}{2M_j}+V(\{\vec q_{\ell}\}), f(\alpha,\vec q_i)\Bigg]
\cr =& \Bigg[\frac {\vec p_i^{\,2}}{2M_i}, f(\alpha,\vec
q_i)\Bigg] \cr =& \frac{-i}{2M_i}[\vec p_i \cdot \vec \nabla_{\vec
q_i} f(\alpha,\vec q_i)+\nabla_{\vec q_i} f(\alpha,\vec q_i) \cdot
\vec p_i]~~~,
\end{align}
giving
\begin{equation}\label{eq:comm2}
[[H,f(\alpha,\vec q_i)],f(\alpha,\vec
q_j)]=-\delta_{ij}\frac{1}{M_i}[\vec \nabla_{\vec q_i} f(\alpha,\vec
q_i)]^2~~~.
\end{equation}
Substituting this into \eqref{eq:enprod2}, we obtain finally
\begin{equation}\label{eq:enprod3}
 \frac{d}{dt}{\rm
Tr}H\rho(t)=\gamma \sum_{\alpha}{\cal F}(\alpha,t)\sum_i\frac
{m_i^2}{M_i} {\rm Tr}\big( [\vec \nabla_{\vec q_i} f(\alpha,\vec
q_i)]^2\rho(t) \big)~~~.
\end{equation}

A further simplification of this result can be achieved by using
the Fourier transform representation of $F(\vec x-\vec y,t)$,
which (recalling the assumed spatial inversion invariance) takes
the form
\begin{align}\label{eq:fourrep}
F(\vec x-\vec y,t)=&\int \frac {d^3k} {(2 \pi)^3} \cos\Big( \vec k
\cdot (\vec x-\vec y\,) \Big) \hat F(\vec k, t)\cr =&\int \frac
{d^3k} {(2 \pi)^3}[\cos(\vec k \cdot \vec x) \cos (\vec k \cdot
\vec y) + \sin(\vec k \cdot \vec x) \sin( \vec k \cdot \vec y)
]\hat F(\vec k, t)~~~.\cr
\end{align}
This has the general structure of \eqref{eq:factor}, with
$\sum_{\alpha}$ corresponding to $\int d^3 k (2\pi)^{-3}
\sum_{n=1}^2$, with $n$ a discrete index distinguishing between
the sine and cosine modes, that is, ${\cal F}(\alpha,t)=\hat
F(\vec k,t)$ for both $n=0,1$, and  $f(\vec k, n=0,\vec
x)=\cos(\vec k \cdot \vec x)$ and $f(\vec k, n=1,\vec x)=\sin(\vec
k \cdot \vec x)$.  Substituting \eqref{eq:fourrep} into
\eqref{eq:enprod3} then gives
\begin{align}\label{eq:enprod4}
\frac{d}{dt}{\rm Tr}H\rho(t)=&\gamma \int \frac {d^3 k}{(2\pi)^3}
\hat F(\vec k,t) \sum_i\frac {m_i^2}{M_i} {\rm Tr}\Big[ \vec
k^{\,2} \Big(\cos^2(\vec k\cdot \vec q_i) + \sin^2 (\vec k \cdot
\vec q_i) \Big) \rho(t) \Big]\cr =&\gamma  \int \frac {d^3
k}{(2\pi)^3} \vec k^{\,2}  \hat F(\vec k,t) \sum_i\frac
{m_i^2}{M_i} ~~~,\cr
\end{align}
where in the final step we have used ${\rm Tr}\rho(t)=1$.  Thus,
in the Markovian approximation, we get a simple formula for the
energy production rate, expressed entirely in terms of the Fourier
transform of $F(\vec x-\vec y,t)$.  We see that the dynamics of
the density matrix $\rho(t)$ drops out  of the final formula, as
does the interaction potential in the Hamiltonian $H$, leaving a
result that is just the sum of contributions from the kinetic
terms of the individual particles.

We give now several specific examples of the formula
\eqref{eq:enprod4}.  First of all, in the standard white noise CSL
model, one has uncoupled space and time correlators of the product
form
\begin{equation}\label{eq:factorcorr}
D(\vec x-\vec y,t-s)=G(\vec x-\vec y) \delta(t-s)~~~,
\end{equation}
 which taking account of the fact that $\int_0^t ds
\delta(t-s)=1/2$, gives
\begin{equation} \label{eq:fwhite}
F(\vec x-\vec y,t)=\frac{1}{2}G(\vec x-\vec y\,)~~~.
\end{equation}
The spatial correlation function $G(\vec x-\vec y\,)$ is the
autoconvolution of the function $g(\vec x)$ introduced as the CSL
smearing function,
\begin{align}\label{eq:convol}
G(\vec x-\vec y\,)=&\int d^3z g(\vec x-\vec z)g(\vec y-\vec
z)~~~,\cr g(\vec x)=&\Bigg( \sqrt{2\pi} r_C\Bigg)^{-3} e^{-\vec
x^2/(2 r_C^2)}~~~,\cr
\end{align}
which, incidentally, gives an alternative form of factor
decomposition for this model.  We will continue, however, to use
the factor decomposition given by the Fourier transform, which is
\begin{equation}\label{eq:fourg}
G(\vec x-\vec y\,)=\int \frac {d^3 k} {(2\pi)^3} \cos\Big( \vec k
\cdot (\vec x -\vec y\,)\Big) e^{-\vec k^2 r_C^2}~~~,
\end{equation}
so that substituting \eqref{eq:fourg}   and  \eqref{eq:fwhite}
into \eqref{eq:fourrep} and writing $k=|\vec k|$ we have
\begin{equation}\label{eq:fourf}
\hat F(\vec k,t) =\frac{1}{2}e^{-k^2r_c^2}~~~.
\end{equation}
Substituting this into \eqref{eq:enprod4}  gives for the white
noise CSL model
\begin{align}\label{eq:enprod5}
\frac{d}{dt}{\rm Tr}H\rho(t)=&\frac {\gamma}{4\pi^2}  \sum_i\frac
{m_i^2}{ M_i} \int_0^{\infty} dk k^4 e^{-k^2 r_C^2}\cr =& \frac {3
\gamma}{32 \pi^{3/2}r_C^5} \sum_i\frac {m_i^2}{ M_i} = \frac {3
\lambda}{4 m_N^2 r_C^2} \sum_i\frac {m_i^2}{ M_i}~~~,\cr
\end{align}
with $m_N$ the nucleon mass and $\lambda=\gamma m_N^2 /(8 \pi^{3/2}
r_C^3)$ the alternative form of the noise coupling generally used in
the CSL literature.\footnote{ In the CSL model literature, what we
here call $\gamma m_N^2$ is termed $\gamma$, because the noise there
is introduced as coupled to the nucleon number density rather than
the mass density.   Also, we note that the dimensionality of
$\gamma$ is determined by the dimensionality assigned to the field
$\phi$, and is not the same in our white noise and thermal model
examples. In the white noise CSL model, what we call $\gamma$ has
dimensionality ${\rm mass}^{-4}$ in microscopic units with
$\hbar=c=1$, whereas in the thermal noise model discussed below,
where $\phi$ is taken as a conventional dimension one boson field,
$\gamma$ has dimensionality ${\rm mass}^{-2}$ in microscopic units.}
This result agrees with the standard answer for the constant energy
production rate in the CSL model.

Consider next a variant of the product correlator model, in which
there is a cutoff in the frequency spectrum, obtained by replacing
$\gamma \delta(t-s)$ in \eqref{eq:factorcorr} by
\begin{equation}\label{eq:delgam}
\delta_{\gamma(\omega)}(t-s)=\frac{1}{\pi} \int_0^{\infty} d\omega
\gamma(\omega) \cos \Big( \omega (t-s)\Big )~~~.
\end{equation}  
This replacement turns the original coupling $\gamma$  into a 
frequency dependent coupling $\gamma(\omega)$, with the specialization back to constant $\gamma$  given by  $\delta_{\gamma(\omega)\equiv \gamma}=\gamma\delta(t-s)$. In this case we
find that
\begin{equation}\label{eq:deltomint}
\int_0^t ds \delta_{\gamma(\omega)}(t-s)= \frac {1}{\pi}
\int_0^{\infty} \gamma\Big(\frac{u}{t}\Big)\frac {du}{u} \sin
u~~~,
\end{equation}
which approaches the constant $\gamma(0)/2$ as $t\to \infty$, with
the entire contribution in the infinite time limit coming from the
infrared  region of the integral near $\omega=0$.  Thus even with
a high frequency cutoff, there is a constant energy production
rate at large times in a model with uncoupled space and time
correlators.  To avoid getting a constant energy production rate
in the product correlator model,  one must include an infrared
cutoff, by taking $\gamma(0)=0$.

Finally, anticipating our  discussion below of thermal noise,
consider a correlator of the general form
\begin{equation}\label{eq:thermd}
D(\vec x-\vec y,t-s)=\int \frac {d^3k}{(2\pi)^3 \omega_k} f(k)
\cos \Big(\vec k \cdot (\vec x- \vec y\,) \Big) \cos \Big(\omega_k
(t-s)\Big)~~~,
\end{equation}
with $\omega_k$ a wave number dependent angular frequency.
Integrating to form $F(\vec x-\vec y, t)$, we have
\begin{equation}\label{eq:thermf}
F(\vec x-\vec y,t) = \int_0^t ds D(\vec x-\vec y,t-s) =\int \frac
{d^3 k}{(2\pi)^3 \omega_k^2} f(k)\cos \Big(\vec k \cdot (\vec x-
\vec y\,) \Big) \sin (\omega_k t)~~~,
\end{equation}
which identifies the Fourier transform as
\begin{equation}\label{eq:fourtherm}
\hat F(\vec k,t)=\frac {f(k)}{\omega_k^2} \sin(\omega_k t)~~~.
\end{equation}
Substituting this into \eqref{eq:enprod4} gives for the energy
production rate
\begin{equation}\label{eq:enprod6}
\frac {d}{dt} {\rm Tr} H\rho(t)=\gamma \sum_i\frac {m_i^2}{M_i}
\int \frac {d^3 k}{(2\pi)^3 } \frac {k^2 f(k)}{\omega_k^2} \sin(
\omega_k t)~~~.
\end{equation}
Even when $\omega_k \propto k$, this expression is  strongly
convergent in the infrared as a  consequence of the vanishing of
phase space for small $k$ values. Hence if $f(k)$ is cut off sharply
at large $k$ values, as expected in thermal models, it leads to a
vanishing energy production rate at large times by use of the
Riemann-Lebesgue theorem.  Integrating to find the total energy
production $\Delta {\rm Tr} H \rho(t) \equiv {\rm Tr} H \rho(t) -
{\rm Tr} H \rho(0)$, we find
\begin{equation}\label{eq:enprodtot}
\Delta {\rm Tr} H \rho(t)=\frac{\gamma}{2\pi^2}  \sum_i\frac
{m_i^2}{M_i} \int_0^{\infty} dk \frac {k^4 f(k)}{\omega_k^3} [
1-\cos( \omega_k t)]~~~,
\end{equation}
which  as $t \to \infty$ gives, again by an application of the
Riemann-Lebesgue theorem,
\begin{equation} \label{eq:enprodasympt}
\Delta {\rm Tr} H \rho(\infty)=\frac{\gamma}{2\pi^2}  \sum_i\frac
{m_i^2}{M_i} \int_0^{\infty} dk \frac {k^4 f(k)}{\omega_k^3}~~~.
\end{equation}

\subsection{Gamma radiation from atoms}

An important constraint on noise model parameters is provided by
the spontaneous emission of gamma rays from atoms, a process first
calculated for free electrons by Fu ~\cite{fu} and later
calculated for general atomic systems by Adler and Ramazano\v glu
~\cite{fethi}.  Results were given in the latter paper for a
correlator of the form $G(\vec x-\vec y\,)
\delta_{\gamma(\omega)}(t-s)$. Remembering that the CSL definition
of $\gamma$ is $m_N^2$ times the definition of $\gamma$ used in
this paper, and comparing \eqref{eq:fwhite} and \eqref{eq:fourg}
with \eqref{eq:thermd}, we see that  the results of ~\cite{fethi}
can be converted to apply to a correlator of the form of
\eqref{eq:thermf} by the substitution
\begin{equation}\label{eq:subs}
\frac{\gamma(\omega)}{m_N^2} e^{-k^2r_C^2} \rightarrow \gamma \pi
\frac{f(k)}{\omega_k} \delta(\omega-\omega_k)~~~.
\end{equation}
When $\omega_k$ has the form $\omega_k=\sqrt{k^2+\mu^2}$, making
this substitution into (44) of \cite{fethi} gives as the
formula for the power radiation $dP$ per unit photon energy $dp$
from a hydrogen atom,
\begin{equation}\label{eq:hpower}
\frac {dP}{dp}= 2\Bigg[1-\frac{1}{[1+(p a_0/2)^2]^2 }\Bigg]  \frac
{\gamma e^2 (p^2-\mu^2)^{3/2} f(\sqrt{p^2-\mu^2})}{3 \pi^2 p}~~~,
\end{equation}
with $e^2 \simeq 1/137.04$ and with $a_0=1/(e^2 m_{\rm
electron})=0.529 \times 10^{-8} {\rm cm}$ the Bohr radius.

\section{Models for the correlation function}
\label{sec:six}

We turn in this section to a discussion of specific models for the
correlation function $D(\vec x -\vec y,t_1-t_2)$ introduced in
\eqref{eq:cm3}. We first briefly consider the standard CSL
factorizable correlation function with white noise, and its
variant with a cutoff in the noise spectrum, which has been the
basis of most discussions to date of the phenomenology of
objective reduction models.  However, one would in general expect
the spatial and temporal structures of the correlation function to
be intertwined, and in particular,   a correlation function
arising from fields with a particle interpretation will have the
spatial and temporal correlations coupled by a mass-shell
constraint. This is the motivation for the models discussed in the
remainder of this section, which are based on a classical model
extracted from the quantum thermal Green's function of a boson of
mass $\mu$.

\subsection{The product correlator model}

The product model for the correlation function was written down
above in \eqref{eq:factorcorr} through \eqref{eq:deltomint}. With a
white noise spectrum, the standard CSL choice for the noise strength
parameter is $m_N^2 \gamma = 10^{-30} {\rm cm}^3 {\rm s}^{-1}$, and
the standard choice for the correlation length is $r_C \sim 10^{-5}
{\rm cm}$.  The white noise model with these parameter choices obeys
all experimental upper bound constraints, and readily explains
measurements in which $n_{\rm out} \sim 10^{13}$ nucleons are
displaced by a distance of at least $r_C$.

In \cite{adler} Adler gave a reanalysis of the upper and lower
bounds on parameters in stochastic reduction models.  Under the
assumption that latent image formation, in either photography or
etched track detectors, constitutes a measurement (rather than the
measurement occurring only through the subsequent development that
reveals the latent image), he concluded that the noise strength
parameter $\gamma$ should be larger than conventionally assumed in
the CSL white noise model, by a factor of $2 \times 10^{9\pm 2}$.
This however conflicts with bounds set by Fu \cite{fu} and Adler
and Ramazano\v glu \cite{fethi} on spontaneous 11 keV gamma
radiation emission from germanium, unless the white noise spectrum
is cut off at energies below 11 keV by the spectral weight
$\gamma(\omega)$ appearing in \eqref{eq:delgam}.  Such a cutoff
would still allow sufficiently rapid state vector reduction to
account for observed measurement times, as already noted in the
review of Bassi and Ghirardi \cite{bgrev}.  Thus, the product
model for the correlation function, with a high frequency cutoff
in the noise spectrum, is consistent both with all upper bounds,
and with the assumption that latent image formation constitutes a
measurement signaling state vector reduction.  Such a correlation
function might arise from a pre-quantum theory in which quantized
fields are not the primary entities, as in \cite{adlbook}.  But as
already noted, a product correlation function is not expected to
arise from quantum fields with a particle interpretation.

\subsection{Thermal correlation function model}

In this section we shall motivate a model for the correlation
function $D(\vec x-\vec y, t_1-t_2)$ by considering the
correlation function for a quantum field in a thermal state at
temperature $T$.  Let $\phi(\vec x, t)$ be a scalar quantum field,
with the mode decomposition
\begin{equation}\label{eq:scalfield}
\phi(\vec x, t)=\int d^3 k \Big[ \frac{1}{2\omega_k
(2\pi)^3}\Big]^{1/2}\Big[a(\vec k) e^{i(\vec k \cdot \vec x -
\omega_k t)} + a^{\dagger}(\vec k) e^{-i(\vec k \cdot \vec x -
\omega_k t)} \Big]~~~,
\end{equation}
where $a(\vec k)$ and $a^{\dagger}(\vec k)$ are the mode
annihilation and creation operators, and where the mode energy
$\omega_k$ is
\begin{equation}\label{eq:omegdef}
\omega_k=\sqrt{\vec k^2 +\mu^2}~~~,
\end{equation}
with $\mu$ the scalar field mass. We have, as before, set Planck's
constant $\hbar$ equal to unity, and also set the Boltzmann
constant equal to unity, so that in a thermal state at temperature
$T$ the expectations of products of creation and annihilation
operators are given by
\begin{align}\label{eq:prodexpect}
\langle a(\vec k) a^{\dagger}(\vec k^{\prime}) \rangle =&
\delta^3(\vec k -\vec k^{\prime}) [1+N(\vec k)]~~~,\cr \langle
a^{\dagger}(\vec k)  a(\vec k^{\prime}) \rangle =& \delta^3(\vec k
-\vec k^{\prime}) N(\vec k)~~~,\cr
\end{align}
with the mean occupation number $N(\vec k)$ given by
\begin{equation}\label{eq:ndef}
N(\vec k)=\frac{1} {e^{\frac {\omega_k}{T} }-1 }~~~.
\end{equation}
{}From these equations we can now calculate the correlation function
\begin{equation}\label{eq:corrfn}
\langle \phi(\vec x, t_1)\phi(\vec y, t_2)\rangle=\int d^3 k
\frac{1}{2\omega_k (2\pi)^3} \Big[[1+N(\vec k) ] e^{i\big(\vec k
\cdot (\vec x -\vec y)- \omega_k (t_1-t_2)\big)} + N(\vec k)
e^{-i\big(\vec k \cdot (\vec x-\vec y) - \omega_k (t_1-t_2)\big)}
\Big]~~~,
\end{equation}
which can be written as the sum of a temperature-independent part
$\Delta_+(\vec x-\vec y,t_1-t_2)$ and a temperature-dependent part
$D(\vec x-\vec y,t_1-t_2)$, as follows,
\begin{align}\label{eq:corrsplit}
\langle \phi(\vec x, t_1)\phi(\vec y, t_2)\rangle=&\Delta_+(\vec
x-\vec y,t_1-t_2) + D(\vec x-\vec y,t_1-t_2)~~~,\cr \Delta_+(\vec
x-\vec y,t_1-t_2)=&\int d^3 k \frac{1}{2\omega_k (2\pi)^3}
e^{i\big(\vec k \cdot (\vec x -\vec y\,)- \omega_k
(t_1-t_2)\big)}~~~,\cr D(\vec x-\vec y,t_1-t_2)=&\int d^3 k
\frac{N(\vec k)}{2\omega_k (2\pi)^3} \Big[e^{i\big(\vec k \cdot
(\vec x -\vec y\,)- \omega_k (t_1-t_2)\big)} +  e^{-i\big(\vec k
\cdot (\vec x-\vec y\,) - \omega_k (t_1-t_2)\big)} \Big]~~~.\cr
\end{align}
In the zero temperature limit, $D(\vec x,t)$ vanishes, and
\eqref{eq:corrsplit} reduces to the temperature-independent piece
$\Delta_+$, which is one of the standard relativistic quantum
theory vacuum Green's functions arising directly from the
non-commutativity of $a(\vec k)$ and $a^{\dagger}(\vec k)$, and is
a complex number for general arguments.  The real-valued
temperature-dependent piece $D(\vec x-\vec y,t_1-t_2)$, on the
other hand,  is invariant under the interchange $\vec x,t_1
\leftrightarrow \vec y,t_2$, and therefore can serve as a model
for the expectation of  real, classical, commuting noise fields
introduced in \eqref{eq:cm3}.

Since $N(\vec k)$ and $\omega_k$  are even in $\vec k$, writing $
e^{\pm i\big(\vec k \cdot (\vec x -\vec y\,)\big)}=\cos\big(\vec k
\cdot (\vec x -\vec y\,)\big)\pm i \sin \big(\vec k \cdot (\vec x
-\vec y\,)\big)$, the sine functions average to zero, and the
formula for $D(\vec x-\vec y,t_1-t_2)$ simplifies to
\begin{equation}\label{eq:dsimp}
 D(\vec x-\vec y,t_1-t_2)=\int\frac{ d^3 k}{(2\pi)^3}
\frac{N(\vec k)}{\omega_k } \cos\big(\vec k \cdot (\vec x -\vec
y\,)\big)\cos\big(\omega_k (t_1-t_2)\big)~~~,
\end{equation}
which has the form assumed in \eqref{eq:thermd}, with $f(k)=N(\vec
k)$ as given in \eqref{eq:ndef}  and with $\omega_k$ the
energy-momentum relation given in \eqref{eq:omegdef}.  We shall
slightly generalize the model specified by \eqref{eq:dsimp} and
\eqref{eq:ndef}, by introducing a thermodynamic chemical potential $\zeta$ into
the occupation number, which we thus write as
\begin{equation}\label{eq:ndef1}
N(\vec k)=\frac{1} {e^{\frac {\omega_k-\zeta }{T}}-1 }~~~,
\end{equation}
which allows us to accommodate systems with general particle density \cite{kadbaym}.  
For the case of noise fields associated with particles having a
standard energy-momentum dispersion relation, \eqref{eq:dsimp},
\eqref{eq:omegdef}, and \eqref{eq:ndef1} constitute our basic
model for the correlation function $D(\vec x-\vec y,t_1-t_2)$.
Corresponding to this model, the function $F(\vec x-\vec y,t)$
defined in \eqref{eq:fdeff} is given by
\begin{equation}\label{eq:fintegral}
F(\vec x-\vec y,t)=\int_0^tds D(\vec x-\vec y,t-s) =\int\frac
 {d^3 k}{(2\pi)^3}
\frac{N(\vec k)}{\omega_k^2 } \cos\big(\vec k \cdot (\vec x -\vec
y\,)\big)\sin\big(\omega_k t\big)~~~,
\end{equation}
and the integral appearing in the  rate function $\Gamma(t)$ of
\eqref{eq:gamdef} is given by
\begin{equation}\label{eq:redint}
I(\vec x-\vec y,t)\equiv \int_0^t ds F(\vec x-\vec y,s)
 =\int \frac
 { d^3 k}{(2\pi)^3}
\frac{N(\vec k)}{\omega_k^3 } \cos\big(\vec k \cdot (\vec x -\vec
y\,)\big)\big[1-\cos\big(\omega_k t\big)\big]~~~.
\end{equation}

\subsection{Dilute and nonrelativistic limits}

Let us consider now the dilute limit of \eqref{eq:dsimp} and
\eqref{eq:ndef1}, obtained \cite{kadbaym} by letting  the chemical
potential $\zeta$ be large and negative, so that $N(\vec k)$
becomes
\begin{equation}\label{eq:ndef2}
N(\vec k) \simeq e^{-\frac {\omega_k-\zeta}{T} }~~~.
\end{equation}
We will be particularly interested in applying \eqref{eq:ndef2} to
the nonrelativistic case $T<<\mu$, where we can expand
\begin{equation}\label{eq:expand}
\omega_k=\sqrt{\vec k^2 +\mu^2}\simeq \mu + \frac {\vec
k^2}{2\mu}~~~,
\end{equation}
so that $N(\vec k)$ becomes
\begin{equation}\label{eq:ndef3}
N(\vec k)=e^{-(\mu-\zeta)/T} e^{-\vec k^2/(2\mu T)}~~~.
\end{equation}
Where $\omega_k$ appears as a denominator factor  in
\eqref{eq:dsimp}, \eqref{eq:fintegral}, and \eqref{eq:redint}, it
can be approximated by $\mu$, so these equations become
respectively
\begin{align}\label{eq:muapprox}
 D(\vec x,t)\simeq& \frac{ e^{-(\mu-\zeta)/T}}{\mu} \int\frac{ d^3 k}{(2\pi)^3}
  e^{-\vec k^2/(2\mu T)}~\cos\big(\vec k \cdot \vec x \big)\cos\big((\mu+\frac {\vec k^2}{2\mu})t\big)~~~,\cr
F(\vec x,t)\simeq& \frac{ e^{-(\mu-\zeta)/T}}{\mu^2} \int\frac{
d^3 k}{(2\pi)^3}
  e^{-\vec k^2/(2\mu T)}~\cos\big(\vec k \cdot \vec x \big)\sin\big((\mu+\frac {\vec k^2}{2\mu})t\big)~~~, \cr
I(\vec x,t)\simeq& \frac{ e^{-(\mu-\zeta)/T}}{\mu^3} \int\frac{
d^3 k}{(2\pi)^3}
  e^{-\vec k^2/(2\mu T)}~\cos\big(\vec k \cdot \vec x \big)\big[1-\cos\big((\mu+\frac {\vec
k^2}{2\mu})t\big)\big]~~~.\cr
\end{align}
Carrying out the angular averaging over $\vec k$,  remembering
that it is the difference $D(\vec 0,t)-D(\vec x,t)$ that enters
into the reduction formalism, and writing $R=|\vec x|$, $k=|\vec
k|$, \eqref{eq:muapprox} yields
\begin{align}\label{eq:muapprox1}
D(\vec 0,t)- D(\vec x,t)\simeq& \frac{e^{-(\mu-\zeta)/T}}{\mu}
\int_{-\infty}^{\infty}\frac{k^2dk}{(2\pi)^2}
  e^{- k^2/(2\mu T)}\Big[1-\frac{\sin(kR)}{kR}\Big] \cos\big((\mu+\frac { k^2}{2\mu})t\big)~~~,\cr
F(\vec 0,t)-F(\vec x,t)\simeq& \frac{ e^{-(\mu-\zeta)/T}}{\mu^2}
\int_{-\infty}^{\infty}\frac{k^2dk}{(2\pi)^2}
  e^{- k^2/(2\mu T)} \Big[1-\frac{\sin(kR)}{kR}\Big]  \sin\big((\mu+\frac { k^2}{2\mu})t\big)~~~, \cr
I(\vec 0,t)-I(\vec x,t)\simeq& \frac{ e^{-(\mu-\zeta)/T}}{\mu^3}
\int_{-\infty}^{\infty}\frac{k^2dk}{(2\pi)^2}
  e^{- k^2/(2\mu T)} \Big[1-\frac{\sin(kR)}{kR}\Big]  \big[1-\cos\big((\mu+\frac {
k^2}{2\mu})t\big)\big]~~~.\cr
\end{align}

The integrals in \eqref{eq:muapprox1} can all be evaluated from
the  formula
\begin{equation}\label{eq:intform}
\int_{-\infty}^{\infty} x^2dx \exp(-\alpha x^2)\frac{\sin
(x\beta)}{x\beta}=\frac{\sqrt{\pi}}{2
\alpha^{3/2}}e^{-\beta^2/(4\alpha)}~~~,
\end{equation}
with results that are summarized in Appendix B.  In particular,
for large times, the formula for $I(\vec 0,t)-I(\vec x,t)$ limits
to
\begin{equation}\label{eq:larget}
I(\vec 0,t=\infty)-I(\vec x,t=\infty)\simeq
e^{-(\mu-\zeta)/T}\Big(\frac{T}{2\pi\mu}\Big)^{3/2} \Big[1-e^{-
R^2 \mu T/2} \Big]~~~
\end{equation}
and the formulas of Appendix B show that the characteristic
reduction time $t_R$ for approach to the asymptotic value of
\eqref{eq:larget} is the inverse temperature $T^{-1}$.

To compare this to the standard CSL model formulas, let us look at
the decay of the off-diagonal density  matrix element of a one
particle system of mass equal to the nucleon mass $m_N$, which we
have seen is governed by the same rate function $\Gamma$ as state
vector reduction.  From \eqref{eq:densdecay} and
\eqref{eq:gamnew}, at large times one has
\begin{equation}\label{eq:tlarge1}
\langle \vec r^{\,1}|\rho(t=\infty)|\vec r^{\,2}\rangle =
e^{-\Gamma(\infty)}\langle \vec r^{\,1}|\rho(0)|\vec
r^{\,2}\rangle~~~,
\end{equation}
with
\begin{align}\label{eq:gaminf}
\Gamma(t=\infty)=&2\gamma m_N^2 [I(\vec 0,\infty)-I(\vec
r^{\,1}-\vec r^{\,2},\infty)]\cr =& 2\gamma
m_N^2e^{-(\mu-\zeta)/T}\Big(\frac{T}{2\pi\mu}\Big)^{3/2}
\Big[1-e^{- R^2 \mu T/2} \Big]~~~,\cr
\end{align}
where we have written $R=|\vec r^{\,1}-\vec r^{\,2}|$. The
comparable formula in the CSL model is given in
 (8.15) of \cite{bgrev},
\begin{equation}\label{eq:rhocsl}
\langle \vec r^{\,1}|\rho(t)|\vec r^{\,2}\rangle = e^{-\Gamma^{\rm
CSL}(t)}\langle \vec r^{\,1}|\rho(0)|\vec r^{\,2}\rangle~~~,
\end{equation}
where $\Gamma^{\rm CSL}(t)$ is given by
\begin{equation}\label{eq:gamcsl}
\Gamma^{\rm CSL}(t)=t  \gamma^{\rm CSL} \Big( \frac{1}{4\pi
r_c^2}\Big)^{3/2} \Big[1-e^{-R^2/(4r_c^2)}\Big]~~~,
\end{equation}
and where (as remarked above in a footnote) $\gamma^{\rm CSL}$ is
what we call $\gamma m_N^2$. We see that the functional form of the
$R$-dependence in \eqref{eq:larget} and \eqref{eq:gamcsl} is the
same, with the CSL model correlation length $r_C$ related to the
nonrelativistic thermal model parameters by
\begin{equation}\label{eq:nonrelrc}
r_C^2=\frac{1}{2\mu
T}~~~,~~~\Big(\frac{T}{2\pi\mu}\Big)^{3/2}=\mu^{-3}\Big(
\frac{1}{4\pi r_C^2}\Big)^{3/2}~~~.
\end{equation}
However, whereas $\Gamma^{\rm CSL}(t)$ grows linearly with time
for large times $t$, in the thermal noise model $\Gamma(t=\infty)$
approaches a constant.  This means that to achieve the degree of
density matrix diagonalization, or state vector reduction,
attained  in the CSL model in time $\Delta t$, the parameters in
the thermal model must obey
\begin{equation}\label{eq:paramval}
\Delta t   \gamma^{\rm CSL}= \frac{2\gamma
m_N^2}{\mu^3}e^{-(\mu-\zeta)/T}~~~.
\end{equation}

\subsection{Can thermalized dark matter be the noise source?}

As we have already noted, one motivation for studying non-white
noise is to investigate whether there can be a cosmological origin
for the noise that drives state vector reduction in objective
reduction models.  Since there is now strong evidence that about a
quarter of the closure density of the universe consists of dark
matter, and since weakly interacting massive particle (WIMP)
candidates for dark matter are expected to be thermalized, it is
natural to apply the results of the preceding section to an
analysis of whether dark matter can account for the noise coupling
in \eqref{eq:cm4}.  We will not attempt to discuss here the
necessary conditions for dark matter to give a real-valued, as
opposed to an imaginary-valued, noise term in the Schr\"odinger
equation; this important question will be deferred to future work.
What we shall do in this section is to assume that a real-valued
noise coupling can be achieved, and to investigate the
phenomenological implications of assuming that state vector
reduction is associated with observed dark matter parameters.

A few basic facts about dark matter are needed.   If dark matter
is due to WIMPs, then observational evidence \cite{fiorucci}
suggests a WIMP distribution in the galactic halo of mass density
$\rho_{\rm mass}= 0.3$ GeV ${\rm cm}^{-3}$, and a Maxwellian
velocity distribution with $v_{\rm rms}=220 {\rm km}\, {\rm
s}^{-1}=7.3\times 10^{-4}c$. The r.m.s. velocity is estimated from
the formula
\begin{equation}\label{eq:binding}
\mu v^2/r_{\rm galaxy}=GM_{\rm galaxy}\mu/r_{\rm galaxy}^2~~~,
\end{equation}
 which describes the gravitational binding of WIMPs of mass
$\mu$ to the galaxy of mass $M_{\rm galaxy}$, at radius $r_{\rm
galaxy}$, with $G$ the Newton gravitational constant.  Direct
limits on possible solar system-bound dark matter are weaker \cite{frere},
\cite{khrip} by a factor of $3 \times 10^5$, that is, $\rho_{\rm
mass\,\,ss} \leq 0.9 \times 10^5$ GeV ${\rm cm}^{-3}$.  There is
at present no observational limit on possible earth-bound dark
matter.  If there were solar system-bound dark matter, around the
radius of the earth's orbit the r.m.s. velocity, by
\eqref{eq:binding} would be $v_{\rm rms} \sim 30 {\rm km}\,{\rm
s}^{-1}=10^{-4}c$, and for earth-bound dark matter, at the radius
of the earth's surface, the r.m.s. velocity would be $v_{\rm rms}
\sim 8{\rm km}\,{\rm s}^{-1}=0.27\times 10^{-4}c$.

Because the WIMP mass $\mu$ cancels out of \eqref{eq:binding},
there is currently no direct information about the dark matter
particle mass.  Dark matter particles coupling to the mass density
cannot be too light, or they would conflict with gravitational
fifth force experiments.  If we write the noise coupling as
\begin{equation}\label{gammafromM}
\gamma=\frac{1}{M^2}~~~,
\end{equation}
then the fifth force experiments require
\begin{equation}\label{fifth1}
\frac{\exp{(-\mu/\mu_5)}}{M^2} \leq \frac{1}{M_{\rm Planck}^2}~~~,
\end{equation}
with $\mu_5$ the fifth force scale limit, currently \cite{adel} around $\mu_5
\sim 1.4 \times 10^{-3}$eV.
This  gives the lower bound on $M$,
\begin{equation}\label{fifth2}
M\geq 10^{19-0.22 \mu/\mu_5} {\rm GeV}~~~.
\end{equation}
 In addition to this constraint, there
are also model-dependent astrophysical limits on the dark matter
mass; for example, warm dark matter candidates must have masses
greater than 1 keV \cite{viel}.

For a Maxwellian distribution with $N(\vec k)$ given by
\eqref{eq:ndef3}, the r.m.s. velocity is given by
\begin{equation}\label{eq:vrms}
v_{\rm rms}^2=\frac {3 T}{\mu}~~~,
\end{equation}
so that using \eqref{eq:nonrelrc} we have
\begin{equation}\label{eq:vrms1}
v_{\rm rms}=\frac {\sqrt{3/2}}{\mu r_C}~~~.
\end{equation}
Hence for a given dark matter r.m.s. velocity, the correlation
length $r_C$ and the dark matter temperature $T$ are determined as
functions of the dark matter mass $\mu$,
\begin{align}\label{eq:tandrc}
r_C=&\frac{\sqrt{3/2}}{\mu v_{\rm rms}}~~~,\cr T=&\frac{\mu v_{\rm
rms}^2}{3}~~~.\cr
\end{align}
Integrating $N(\vec k)$ over phase space, the number density
$\rho_n$ is given by
\begin{align}\label{eq:numdens}
\rho_n\equiv &\rho_m/\mu=\int \frac{d^3k}{(2\pi)^3} N(\vec k)\cr
 =&\int \frac{d^3k}{(2\pi)^3} e^{-(\mu-\zeta)/T} e^{-\vec k^2 r_C^2}
 =\frac{e^{-(\mu-\zeta)/T}}{8 \pi^{3/2}r_C^3}~~~,\cr
 \end{align}
which determines the factor containing the chemical potential
$\zeta$ in terms of $\rho_m$, $\mu$ and $r_C$,
\begin{equation}\label{eq:chempot}
e^{-(\mu-\zeta)/T}=\frac{\rho_m}{\mu} 8 \pi^{3/2} r_C^3~~~.
\end{equation}

{}From these equations, together with  \eqref{eq:gaminf} and
\eqref{eq:nonrelrc}, and the assumption that the lower bound of
\eqref{eq:lower1} gives a good approximation to the reduction
factor,\footnote{Note, however, that although $\Gamma(t=\infty)$
is positive,  the uniform positivity assumption on the integrand
used to derive the upper and lower bounds is not obeyed in the
thermal model; see the formulas in Appendix B.  Also, the simple model
analyzed in Appendix D gives exponential reduction as in the lower bound, but
with a reduction factor
$e^{-\Gamma(t=\infty)}$, and so the rates calculated from the lower bound
may be optimistic by a factor of two.} we get the
following estimate,
\begin{align}\label{eq:estimate}
{\rm Reduction ~factor} \sim & e^{-2\Gamma(t=\infty)} ~~~,\cr 2
\Gamma(t=\infty) =& 4\Big(\frac{m_N}{M}\Big)^2
\frac{\rho_m}{\mu^4} n^2 N ~~~.\cr
\end{align}
Here, in accordance with the properties of $\Gamma$ discussed in
Sec. 2, $n$ is the number of displaced nucleons that are bunched
within a correlation length $r_C$, and $N$ is the number of such
bunches of displaced nucleons.

Using \eqref{eq:tandrc} and \eqref{eq:estimate}, we can now make
some estimates of the effectiveness of thermal dark matter in
producing state vector reduction in the mass-density coupled
model. Rewriting \eqref{eq:tandrc} in the form
\begin{align}\label{eq:tandrc1}
r_C=&\frac{0.24 \times 10^{-13}}{ v_{\rm rms}}\frac{1 {\rm
GeV}}{\mu}{\rm cm}~~~,\cr T^{-1}=&t_R=\frac{0.2 \times 10^{-23}}
{v_{\rm rms}^2}\frac{\rm 1 GeV}{\mu}{\rm s}~~~,\cr
\end{align}
we get the following tables of values. For the correlation length
$r_C$ in the body of the table in cm, versus the dark matter mass
$\mu$ in keV and its r.m.s. velocity appropriate to the galactic
halo ($v_h=220{\rm km}{\rm s}^{-1}$), solar system-bound dark matter
($v_s=30{\rm km}{\rm s}^{-1}$), and earth-bound dark matter
($v_e=8{\rm km}{\rm s}^{-1}$), we have
\begin{table}[h]
\begin{center}
\begin{tabular}{c|cccccc}
$\mu\to$ & 1 & 10 & $10^2$ & $10^3$ & $10^4$ & $10^6$ \\
\hline $v_h$ & ~~$3\times 10^{-5}$~~ & ~~$3\times 10^{-6}$~~ &
~~$3\times
10^{-7}$~~ & ~~$3\times 10^{-8}$~~ & ~~$3\times 10^{-9}$~~ & ~~$3\times 10^{-11}$~~ \\
$v_s$ & $2\times 10^{-4}$ & $2\times 10^{-5}$ & $2\times
10^{-6}$ & $2\times 10^{-7}$ & $2\times 10^{-8}$ & $2\times 10^{-10}$ \\
$v_e$ & $9\times 10^{-4}$ & $9\times 10^{-5}$ & $9\times 10^{-6}$ &
$9\times 10^{-7}$ & $9\times 10^{-8}$ & $9\times 10^{-10}$
\end{tabular}
\caption{Correlation length $r_C$ (cm) versus $\mu$ (keV) and r.m.s. velocity $v$.}
\end{center}
\end{table}

 Similarly, for the reduction time $t_R$ in seconds in the body of the table, versus the
 dark matter mass and its r.m.s. velocity, we have
\begin{table}[h]
\begin{center}
\begin{tabular}{c|cccccc}
$\mu\to$ & 1 & 10 & $10^2$ & $10^3$ & $10^4$ & $10^6$ \\
\hline $v_h$ & ~~$4\times 10^{-12}$~~ & ~~$4\times 10^{-13}$~~ &
~~$4\times
10^{-14}$~~ & ~~$4\times 10^{-15}$~~ & ~~$4\times 10^{-16}$~~ & ~~$4\times 10^{-18}$~~ \\
$v_s$ & $2\times 10^{-10}$ & $2\times 10^{-11}$ & $2\times
10^{-12}$ & $2\times 10^{-13}$ & $2\times 10^{-14}$ & $2\times 10^{-16}$ \\
$v_e$ & $3\times 10^{-9}$ & $3\times 10^{-10}$ & $3\times 10^{-11}$
& $3\times 10^{-12}$ & $3\times 10^{-13}$ & $3\times 10^{-15}$
\end{tabular}
\caption{Reduction time $t_R$ (s) versus $\mu$ (keV) and r.m.s. velocity v.}
\end{center}
\end{table}

Solving \eqref{eq:estimate} for the value of $\gamma \rho_m$ which
yields $2\Gamma(t=\infty)=1$, which is the minimum value of the exponent
beyond which  reduction of the
state vector starts to occur, we get
\begin{equation}\label{eq:gamrhom}
\gamma \rho_m=\frac{1.5 \times 10^{13}}{n^2N}\Bigg(\frac{\mu}{1{\rm
GeV}}\Bigg)^2 {\rm GeV}{\rm cm}^{-1}~~~.
\end{equation}
{}From this, we get further tables of values.  For $\gamma \rho_m$ in
the body of the table, in ${\rm GeV} {\rm cm}^{-1}$, versus the dark
matter mass $\mu$ in keV, and the effective number of displaced
nucleons $n_{\rm out}=n^2N=10^{22}$ corresponding \cite{bgrev} to
the standard CSL model, or  $n_{\rm out}=n^2N=10^8$ corresponding to
estimates \cite{adler} based\footnote{In the CSL model, one assumes
$n=10^9$, which is the number of nucleons in a volume of linear
dimension $10^{-5} {\rm cm}$, and $N=10^4$, giving $n^2N=10^{22}$.
The latent image estimates of \cite{adler} take $n=5640$ and $N=20$,
giving $n^2N \sim 10^8$.  The CSL model assumes a reduction rate of
$10^7 {\rm s}^{-1}$, whereas the latent image estimates assume a
much smaller reduction rate of 30 ${\rm s}^{-1}$, which is why in a
white noise model the ratio  of the noise strengths between the two
cases is $\sim 10^9$, rather than the ratio $\sim 10^{14}$ of the
$n^2N$ values.} on latent image formation, we have
\begin{table}[h]
\begin{center}
\begin{tabular}{c|cccccc}
$\mu\to$ & 1 & 10 & $10^2$ & $10^3$ & $10^4$ & $10^6$ \\
\hline $10^{22}$ & ~~$2\times 10^{-21}$~~ & ~~$2\times 10^{-19}$~~ &
~~$2\times
10^{-17}$~~ & ~~$2\times 10^{-15}$~~ & ~~$2\times 10^{-13}$~~ & ~~$2\times 10^{-9}$~~ \\
$10^{8}$ & $2\times 10^{-7}$ & $2\times 10^{-5}$ & $2\times 10^{-3}$
& $2\times 10^{-1}$ & $2\times 10^{\phantom{-1}}$ & $2\times 10^{5}$
\end{tabular}
\caption{$\gamma \rho_m$ (GeV ${\rm cm}^{-1})$ versus $\mu$ (keV) and $n_{\rm out}$.}
\end{center}
\end{table}

If we make the assumption that $\gamma =1 ({\rm TeV})^{-2}
=10^{-6}({\rm GeV})^{-1}$, we get a table of values giving
$\rho_m$ in the body of the table, in ${\rm GeV}{\rm cm}^{-3}$,
versus the dark matter mass and the effective number $n_{\rm out}$
of displaced nucleons,
\begin{table}[h]
\begin{center}
\begin{tabular}{c|cccccc}
$\mu\to$ & 1 & 10 & $10^2$ & $10^3$ & $10^4$ & $10^6$ \\
\hline $10^{22}$ & $3$ & ~~$3\times 10^{4}$~~ & ~~$3\times
10^{8}$~~ & ~~$3\times 10^{12}$~~ & ~~$3\times 10^{16}$~~ & ~~$3\times 10^{24}$~~ \\
$10^{8}$ & $~~3\times 10^{14}~~$ & $3\times 10^{18}$ & $3\times
10^{22}$ & $3\times 10^{26}$ & $3\times 10^{30}$ & $3\times 10^{38}$
\end{tabular}
\caption{$\rho_m$ (GeV ${\rm cm}^{-3}$) versus $\mu$ (keV) and $n_{\rm out}$.}
\end{center}
\end{table}

{}From these tables, we see that state vector reduction, by the
standard CSL criterion ($n_{\rm out}=10^{22}$), and with a
correlation length within a decade of the standard CSL value
$r_C=10^{-5} {\rm cm}$, is achievable in the dark matter model for
dark matter masses in the range of 1 to 10 kilovolts, with $\gamma
\sim 1 {\rm TeV}^{-2}$ and with $\rho_m$ below the current upper
limit on solar system-bound dark matter.  Adopting the latent image
criterion ($n_{\rm out}=10^8$) requires dark matter densities that
are much too large, so either the latent image analysis of
\cite{adler} needs modification, or the dark matter model is
unworkable.

For a dark matter mass $\mu$ of a kilovolt or greater, and the
current limit on the fifth force scale $\mu_5$, the fifth force
bound of \eqref{fifth1} becomes
\begin{equation}\label{fifth3}
M\geq 10^{19-0.15\times 10^{6}}~~~,
\end{equation}
which is strongly obeyed for the $M$ values in the GeV to TeV range
that are interesting.  Referring to the discussion following \eqref{eq:dsimp}, and
using \eqref{eq:ndef1} and \eqref{eq:tandrc}, we see that the function $f(\sqrt{p^2-\mu^2})$ in the
formula \eqref{eq:hpower} for the radiated gamma power from a hydrogen atom
becomes
\begin{align}\label{eq:power1}
f(\sqrt{p^2-\mu^2})=&\frac{1}{e^{(p-\zeta)/T}-1}\simeq
e^{-(\mu-\zeta)/T} e^{-(p-\mu)/T}\cr =&\frac
{\rho_m}{\mu}8\pi^{3/2}r_C^3 e^{-3(p-\mu)/(\mu v_{\rm
rms}^2)}~~~.\cr
\end{align}
Since for $\mu$ in the 1 to 10 kilovolt range and for $p=11$ kilovolts, we have
\begin{equation}\label{eq:expon}
\frac{3(p-\mu)}{\mu v_{\rm rms}^2} \geq 6 \times 10^5~~~,
\end{equation}
 the negative exponential in the final factor of \eqref{eq:power1} dominates all other factors in this equation and in \eqref{eq:hpower}, and so the experimental
bound on 11 keV gamma radiation is strongly satisfied.

For both values of $n_{\rm out}$ displayed in the tables, the
reduction time is sufficiently rapid, shorter than a few times
$10^{-9}$ seconds, to account for realizable measurements.   Finally,  the total energy imparted by the noise to an
isolated nucleon is obtained by evaluating \eqref{eq:enprodasympt}
by using the form for $f(k)$ in the dilute nonrelativistic thermal
model, giving
\begin{equation}\label{eq:dilutenprod}
{\rm Tr}H\rho(t=\infty)=\frac{3 m_N \gamma \rho_m}{2 r_C^2
\mu^4}~~~.
\end{equation}
For the CSL value of $n_{\rm out}$, this is smaller than $
10^{-15}$ degrees Kelvin for all values of the dark matter
velocity and mass in the tables, and so is sufficiently small so
as to be unobservable.

The conclusion from this analysis is that, {\it if} dark matter
couplings to ordinary matter have the anti-self-adjoint component
needed  to give a real-valued noise term in the Schr\"odinger
equation, and {\it if} dark matter densities in the vicinity of
earth are larger than the galactic halo density,  but within current
limits on solar system-bound dark matter, one could realize  the
standard CSL reduction model with the standard parameter values,
and obey various important  experimental constraints.  The
italicized assumptions  make this mechanism for realizing state
vector reduction  conjectural; at worst, we have given an
interesting toy model for reduction incorporating a non-white noise
with a mass-shell constraint.

\subsection{Thermal unparticles as the noise source}

Recently Georgi \cite{georgi} has introduced the concept of what he terms
an ``unparticle'', a field characterizing a scale-invariant sector of a low-energy
effective field theory.  This is of interest for collapse models, since if the
noise field of \eqref{eq:cm4} is the low-energy manifestation of a pre-quantum
dynamics, such as discussed in the book \cite{adlbook}, it is
plausible that it could have a scale-invariant structure.  Moreover, such  an unparticle field, if a cosmological relic field, will have  a thermal correlation structure.  The  concept of thermal unparticles has been introduced in a recent paper of Chen et al. \cite {chen}, who construct the thermal unparticle partition function
by using  the observation of Krasnikov \cite{kras}, that an unparticle
field can be constructed as a field with a continuous distribution of mass $\mu^2$, characterized by a scale invariant spectral function $\rho(\mu^2)\propto (\mu^2)^{d-2}$.  More specifically, one obtains the unparticle propagator and
partition function by integrating the corresponding propagator and partition function for a scalar field of squared mass $\mu^2$ over the range $0\leq \mu^2\leq \infty$,
with weighting function $\rho(\mu^2)=(d-1) \Lambda^{2(1-d)} (\mu^2)^{d-2}$. Here
$d$ is the anomalous scaling dimension characterizing unparticle physics, and
$\Lambda$ is a scale parameter (the cutoff for the low-energy effective theory) with
dimension of mass.\footnote{Strictly speaking, the integration over $\mu^2$ should extend only up to $\Lambda^2$, but when the temperature $T<<\Lambda$, the integration for the partition function and thermal correlation function is effectively cut off by $N(\vec k)$ of \eqref{eq:ndef1}, and so negligible error is made in
extending the upper limit to $\infty$.}

In Appendix E we use the same method to construct the unparticle thermal correlation
function from the thermal correlation function of \eqref{eq:dsimp} and \eqref{eq:ndef1}
for a scalar field of mass $\mu^2$.  From this correlation function, we calculate
the integrals needed to study both  the state vector reduction rate and
the noise-induced energy production.  We recapitulate here two key formulas obtained from Appendix E, both of
which apply to a one particle system of mass $m$.  For the decay rate $\Gamma(t)$
of the off-diagonal matrix element $\langle \vec x|\rho(t)|\vec 0\rangle$, which we have seen is also the reduction rate,  we have
\begin{align}\label{eq:decrate}
\Gamma(t)=&2\gamma m^2[I_{\cal U}(\vec 0,t)-I_{\cal U}(\vec x,t)]\cr
=&\frac{\gamma m^2 \Lambda^{2(1-d)} }{\pi^2}
\int_0^{\infty} d\omega \frac{\omega^{2d-3}\big[1-\cos\big(\omega t\big)\big]} {e^{\frac {\omega-\zeta }{T}}-1 }
\int_0^1 dv\big[1-\cos\big(v \omega |\vec x| \big)\big] (1-v^2)^{d-1}~~~,\cr
\end{align}
where the subscript ${\cal U}$ on $I$ corresponds to the notation of \eqref{eq:iint1} of Appendix E.  For the noise-induced energy acquisition rate and total energy
acquired by a particle of mass $m$,
we have from \eqref{eq:enratefinal} and \eqref{eq:enprodfinal} of Appendix E,
\begin{equation}\label{eq:enratefinal1}
\frac{d}{dt}{\rm Tr}H\rho(t)
=\frac{3\gamma m\Lambda^{2(1-d)}}{(2\pi)^2}\frac{\Gamma(3/2)\Gamma(d)}
{\Gamma(3/2+d)}
\int_0^{\infty}d\omega \frac {\omega^{2d} \sin\big(\omega t\big)}
{e^{\frac {\omega-\zeta }{T}}-1 }~~~,
\end{equation}
and
\begin{equation}\label{eq:enrateint}
{\rm Tr}H\rho(t)-{\rm Tr}H\rho(0)
=\frac{3\gamma m\Lambda^{2(1-d)}}{(2\pi)^2}\frac{\Gamma(3/2)\Gamma(d)}
{\Gamma(3/2+d)}
\int_0^{\infty}d\omega \frac {\omega^{2d-1}\big[1- \cos\big(\omega t\big)]}
{e^{\frac {\omega-\zeta }{T}}-1 }~~~.
\end{equation}

Turning our attention first to \eqref{eq:decrate}, we note that the inner integral
over $v$ is always convergent at $v=0$, and is convergent at $v=1$ for ${\rm Re}\, d>0$. Because the inner integral in \eqref{eq:decrate} vanishes as $\omega^2$ for
small $\omega$, the integral over $\omega$ in \eqref{eq:decrate} has precisely the
same convergence properties at $\omega=0$ as the integral giving the total energy
production in \eqref{eq:enrateint}.  To study convergence, there are two cases to consider,
(i) the chemical potential $\zeta$ is negative and nonzero, and (ii) the chemical
potential $\zeta$ is zero.\footnote{The chemical potential must always be less than or
equal to zero, so there is not a third case of positive $\zeta$, which would correspond to a physical region pole in the integrands coming from the vanishing
of the denominator $e^{\frac {\omega-\zeta }{T}}-1 $ in all three integrals.}

In the first case, of strictly negative $\zeta$, the denominator $e^{\frac {\omega-\zeta }{T}}-1 $ is nonzero even at $\omega=0$, and the integrals of
\eqref{eq:decrate} and \eqref{eq:enrateint} converge at $\omega=0$ even when the factor $1- \cos\big(\omega t\big)$ is replaced by unity, as long as ${\rm Re}\, d>0$.  So in this case we can extract the infinite time limit by invoking the
Riemann-Lebesgue theorem, and simply dropping the term $\cos\big(\omega t\big)$ in
\eqref{eq:decrate} and \eqref{eq:enrateint}, giving the formulas
\begin{equation}\label{eq:decrateinfty}
\Gamma(\infty)=\frac{\gamma m^2 \Lambda^{2(1-d)} }{\pi^2}
\int_0^{\infty} d\omega \frac{\omega^{2d-3}} {e^{\frac {\omega-\zeta }{T}}-1 }
\int_0^1 dv\big[1-\cos\big(v \omega |\vec x| \big)\big] (1-v^2)^{d-1}~~~,
\end{equation}
and
\begin{equation}\label{eq:enrateintinfty}
{\rm Tr}H\rho(\infty)-{\rm Tr}H\rho(0)
=\frac{3\gamma m\Lambda^{2(1-d)}}{(2\pi)^2}\frac{\Gamma(3/2)\Gamma(d)}
{\Gamma(3/2+d)}
\int_0^{\infty}d\omega \frac {\omega^{2d-1}}
{e^{\frac {\omega-\zeta }{T}}-1 }~~~.
\end{equation}
Corresponding to the fact that the total energy production is finite, the energy
production rate of \eqref{eq:enratefinal1} vanishes at large time. Referring now
to \eqref{eq:decrateinfty}, we see that there are two subcases governing the
large $|\vec x|$ behavior, which we call
(ia) and (ib).  In subcase (ia), corresponding to ${\rm Re}\, d>1$, the $\omega$
integral is convergent without using the $\omega^2$ factor arising from the inner
integral. So in this subcase we can apply the Riemann-Lebesgue theorem to the
inner integral in the limit of large $|\vec x|$, by dropping the term $\cos\big(v\omega |\vec x|\big)$, leading to the conclusion that $\Gamma(\infty)$ varies from $0$
at $|\vec x|=0$ to a finite value at $|\vec x|=\infty$.  In subcase (ib), corresponding to $1 \geq {\rm Re}\, d >0$, the $\omega^2$ factor from the inner integral is needed for convergence, and on changing integration variable from
$\omega$ to $u=\omega |\vec x|$ one sees that $\Gamma(\infty)$ grows as
$|\vec x|^{2(1-d)}$ as $|\vec x|\to \infty$.

In the second case, of vanishing chemical potential $\zeta$, the denominator $e^{\frac {\omega-\zeta }{T}}-1 $ vanishes at $\omega=0$, and the integrals of
\eqref{eq:decrate} and \eqref{eq:enrateint} now behave for small $\omega$ as
\begin{equation}\label{eq:decratesmall}
\Gamma(t)\sim \frac{\gamma m^2 T\Lambda^{2(1-d)} }{\pi^2}
\int_0 d\omega \omega^{2d-4}\big[1-\cos\big(\omega t\big)\big]
\int_0^1 dv\big[1-\cos\big(v \omega |\vec x| \big)\big] (1-v^2)^{d-1}~~~,
\end{equation}
and
\begin{equation}\label{eq:enrateintsmall}
{\rm Tr}H\rho(t)-{\rm Tr}H\rho(0)
\sim \frac{3\gamma m T\Lambda^{2(1-d)}}{(2\pi)^2}\frac{\Gamma(3/2)\Gamma(d)}
{\Gamma(3/2+d)}
\int_0d\omega \omega^{2d-2}\big[1- \cos\big(\omega t\big)]~~~.
\end{equation}
There are now two subcases, which we label (iia) and (iib).  In subcase (iia)
we have $d>1/2$, and both integrals \eqref{eq:decratesmall} and \eqref{eq:enrateintsmall} converge at $\omega=0$ without using the $\omega^2$ factor
that comes from $1-\cos\big(\omega t\big)$.  So in this case, which behaves much like case (i), we can apply the Riemann-Lebesgue theorem to take the limit as $t \to \infty$ by dropping the term $\cos\big(\omega t\big)$, leading to finite values for   $\Gamma(\infty)$ and ${\rm Tr} H\rho(\infty)-{\rm Tr} H\rho(0)$. One can then
proceed to analyze the large $|\vec x|$ behavior  of $\Gamma(\infty)$, as was done
previously in case (i),  with the
conclusion that this is finite for $d>3/2$ and it behaves as $|\vec x|^{3-2d}$
for $3/2 \geq d>0$.   In subcase (iib), we have $1/2 \geq d> 0$, and the $\omega^2$
coming from the factor $1-\cos\big(\omega t\big)$ is needed for converence;
defining a new integration variable $u=\omega t$, we see that both $\Gamma(t)$
and ${\rm Tr}H\rho(t)$ grow  as $t^{1-2d}$ in the large $t$ limit, and correspondingly, the energy production rate decreases as $t^{-2d}$.  So for
vanishing chemical potential, and $1/2>d>0$, we have the interesting situation that
one achieves perfect reduction at infinite time (that is, $\Gamma(\infty)=\infty$),
although the reduction rate and the total energy production both grow as a fractional power of $t$, rather than linearly
with $t$ as in the standard  CSL model.  Correspondingly, the energy production rate vanishes as a fractional power of $t$ at large time, which should make it easy to
satisfy cosmological constraints \cite{adler} on the noise strength parameter.

We conclude that the thermal unparticle model exhibits a range of interesting behaviors, depending on the values of the chemical potential $\zeta$ and the unparticle dimension $d$.  In addition to these two parameters, the effective noise strength  $\gamma \Lambda^{2(1-d)}$ and the temperature $T$ are also parameters of
the model.  Given the complexities of this four-dimensional parameter space, we do not attempt phenomenological fits of the model to experimental constraints on the noise strength, but this is clearly an interesting topic for future investigation.

\section{Summary and Discussion}

We now summarize what has been done in the preceding sections
and what is in the appendices,
and sketch some directions for extensions of our investigations.  In Secs. I-V we
have continued the study of non-white noise models initiated in (I), focusing on the
special case in which the noise field couples to the particle density.  The analyses  of Secs. II, III, and Appendix D identify the characteristic rate functions governing density matrix diagonalization and state vector reduction, and show that both processes are exponential with the same rate function, in the simplified case
(a single particle in a superposition of two localized states) discussed in Appendix D.  In Sec. IV, we completed our formal analysis for non-white noise by deriving the corresponding Fokker-Planck equation, allowing us to make contact with earlier work
of Pearle \cite{pearle}.  In Sec. V, with an eye towards phenomenological applications,
we analyzed energy production and gamma radiation by atoms in terms of the correlation functions of the non-white noise model.

In Sec. VI we turned to a discussion of specific models for the noise correlation
function.  After a brief discussion of the product correlator model that has been
the basis of most earlier work on objective state vector reduction, we turned to
a detailed analysis of a thermal correlation function model, in which the spatial
and temporal correlations are linked by a mass-shell constraint.  We showed that the
dilute, nonrelativistic limit of the thermal correlator model can be put in
direct correspondence with the formulas of the standard Gaussian CSL model.
We then gave a detailed phenomenological analysis of  thermal dark matter as the noise source, and concluded Sec. VI with a discussion of the behavior of
thermal unparticles as the noise source, sketching qualitative behaviors for
a range of values of the chemical potential and of the unparticle anomalous scaling
dimension. The examples given included cases in which $\Gamma(t)$ and the energy
production both are finite at $t=\infty$, and in which $\Gamma(t)$ and the energy
production both grow as a fractional power smaller than unity as $t \to \infty$.

The appendices deal with various details connected with the main discussion.  In
Appendix A we estimate the validity of the Markovian approximation used in the
energy production discussion, while in Appendix B we compare the master equation
used in our discussion with a more general class of master equations appearing in
the literature.  Appendix C gives the evaluation of integrals for the dilute,
nonrelativistic model, while Appendix E gives details of the unparticle correlation functions.  Appendix D shows that, in a simple model, reduction is
exponential in the rate function $\Gamma(t)$, indicating that the lower bound
derived in Sec. III, as opposed to the upper bound derived there, gives the better estimate of the qualitative reduction behavior.

We can point to a number of possible directions for generalization or extension of
the results of this paper.  (i)  We have considered only the case of a real  noise
coupling, corresponding to an anti-self-adjoint Hamiltonian term.  More generally,
one could consider a complex noise coupling, containing both real and imaginary
noise couplings, with the real term contributing both  to density matrix diagonalization (i.e., decoherence) and to state vector reduction, and the imaginary
part contributing only to decoherence.  (ii) For simplicity, we have only considered a scalar noise
field $\phi$, but a general treatment of non-white noises would allow for the
possibility of spin-1/2 or spin-1 noise fields.  Such an extension may ultimately
be required on phenomenological grounds to make contact with experiment. (iii)
Although we sketched the qualitative behavior of the thermal unparticle case, we
did not attempt to make a quantitative phenomenological survey of the four-dimensional parameter space of this model, and this would be of interest.   (iv)  The derivation of lower
and upper bounds on the reduction rate in Sec. III made use of a positivity assumption, which is not obeyed in the thermal correlator model; can this assumption
be eliminated? (v)  The model calculation of Appendix D  indicated an exponential
dependence of the reduction factor on $\Gamma$, agreeing with the corresponding
density matrix diagonalization calculation but differing by a factor of 2 in the
exponent from the corresponding lower bound of Sec. III.  Can this result be generalized to  the case of many particles and a wave function that is the superposition of many localized states as in \eqref{eq:psisup}?   Clearly, a general argument that the
reduction factor has exponential rather than power law dependence on $\Gamma$
would be significant for the phenomenology of objective reduction models. (vi) In Sec. VIB we formulated our thermal model for the correlation function, by neglecting
the temperature-independent Greens function $\Delta_+$, which reflects the  non-commutativity of creation and annihilation operators. As noted, this gives an effectively classical model for the thermal noise. It would be worth
exploring a fully quantum-mechanical treatment of state vector reduction by a thermal noise field, in which all parts of the quantum mechanical correlation function
\eqref{eq:corrfn} are retained.

\label{sec:seven}

\section{Acknowledgments}

The authors wish to thank  Nima Arkani-Hamed, Lane Hughston, Masataka Fukugita,
Roland Omn\`es, Phillip Pearle, Fethi Ramazano\v glu, and Fransisco
Yndur\'ain, for informative conversations or emails.

The work of SLA was supported in part by the Department of Energy
under grant no DE-FG02-90ER40542. The work of AB was supported in
part by DGF (Germany) and by INFN-Italy. Both SLA and AB wish to
acknowledge the hospitality of Clare Hall in  Cambridge, and AB also
wishes to acknowledge the hospitality of the Institute for Advanced
Study in Princeton, and SLA that of the Aspen Center for Physics.

\appendix\section{Markovian approximation}

One can estimate the validity of the Markovian approximation by
considering the case of a single free particle of mass $m$, so
that $H=p^2/(2m)$.  Then we easily calculate that
\begin{equation}\label{eq:markov1}
M(\vec y,s-t)=m\delta^3\big(\vec y-e^{iH(s-t)}\vec
qe^{-iH(s-t)}\big) =m\delta^3\big(\vec y-\vec q - (\vec
p/m)(s-t)\big)~~~,
\end{equation}
so that repeating the steps leading to \eqref{eq:enprod6} we find
\begin{equation}\label{eq:markov2}
\frac {d} {dt}{\rm Tr}\rho(t) H=  {\gamma m} \int_0^t ds \int
\frac {d^3
k}{(2\pi)^3}\frac{f(k)}{\omega_k}\cos\Big(\omega_k(t-s)\Big) {\rm
Tr}{\cal O}(s-t)~~~,
\end{equation}
with ${\cal O}(s-t)$ given by
\begin{equation}\label{eq:markov3}
{\cal O}(s-t)=-\frac{1}{2}[e^{-i\vec k\cdot \big(\vec q+ (\vec
p/m)(s-t)\big)},[e^{i\vec k \cdot \vec q} ,\vec p^{\,2}]]~~~.
\end{equation}
This expression can be simplified by use of the Baker-Hausdorff
theorem and the canonical commutation relations, giving after
considerable algebra, and dropping terms that are odd in $\vec k$,
\begin{equation}\label{eq:markov4}
{\cal O}(s-t)=  k^2 \cos\Big( \frac{\vec k \cdot \vec p}{m}(s-t)
\Big)\cos \Big(
 \frac {k^2}{2m} (s-t)\Big) - 2\vec p\cdot \vec k   \sin\Big( \frac{\vec k \cdot \vec p}{m}(s-t)
\Big)     \sin \Big(
 \frac {k^2}{2m}(s-t) \Big) ~~~.
 \end{equation}
 We see that all dependence of ${\cal O}(s-t)$ on $s-t$ is through
 oscillatory terms.
 Assuming that the characteristic
 spatial variation scale of the problem is governed by $\omega_k \sim |\vec k | \sim |\vec
 p\,| \sim k_{\rm max}$, with $k_{\rm max}$ the characteristic $k$-value at which
 $f(k)$ cuts off,  then when the particle mass $m$ is large enough for the kinetic energy
 at $k_{\rm max}$ to obey
 \begin{equation}\label{condition}
 \frac{k_{\rm max}^2}{2m} << k_{\rm max}~~~,
 \end{equation}
 the variation of ${\cal O}(s-t)$ with $s$ is much slower than
 that of the cosine factor in \eqref{eq:markov2}.  In this case the
 integral in \eqref{eq:markov2} is well approximated by replacing
 ${\cal O}(s-t)$ by ${\cal O}(0)=k^2$, which recovers the result
 of the Markovian approximation made in Sec. V.

 \section{Comparison with master equations for decoherence}

A further understanding of the effect of the thermal field
$\phi(\vec{x},t)$ on the evolution of the wave function can be
obtained by comparing \eqref{eq:lindnw0} for the density matrix
with typical master equations used for describing open quantum
systems. Here we will follow the path outlined in~\cite{2007-vac},
where a comparison of this kind has been made between the GRW
model~\cite{1986-grw} and collisional
decoherence~\cite{2000-vac,2001-vac,kla}.

We consider the evolution of a single particle; under the
Markovian approximation ($M(\vec{y}, s - t) = M(\vec{y}, 0) =
M(\vec{y}\,)$) discussed in sec. V, \eqref{eq:lindnw0} reads:
\begin{equation}
\frac{d}{dt} \rho(t) \; = \;  -i [H, \rho(t)] + {\mathcal
L}_{t}^{\phi} [ \rho(t) ],
\end{equation}
with
\begin{equation} \label{eq:xcfg}
{\mathcal L}_{t}^{\phi} [ \rho ] \; = \; - \gamma \int d^3 x \int
d^3 y \; [ M(\vec{x}), [ M(\vec{y}), \rho ]] \; F(\vec{x} - \vec{y},
t),
\end{equation}
and $M(\vec{x}) = m \delta^3(\vec{x} - \vec{q})$. The term
${\mathcal L}_{t}^{\phi}$, which includes the effect of the thermal
field $\phi(\vec{x},t)$ on $\rho(t)$, is the one we will focus
 on. Let us introduce the Fourier transform:
\begin{equation}
F(\vec{x} - \vec{y}, t) \; = \; \int \frac{d^3k}{(2\pi)^3}\,
\hat{F}(\vec{k},t) \, e^{i \vec{k} \cdot (\vec{x} - \vec{y}\,)},
\end{equation}
with $\hat{F}(\vec{k},t) = \hat{F}(-\vec{k},t)$ due to spatial
inversion invariance. One can rewrite~\eqref{eq:xcfg} in terms of
$\hat{F}(\vec{k},t)$ as follows:
\begin{equation} \label{eq:me}
{\mathcal L}_{t}^{\phi} [ \rho ] \; = \; 2 m^2 \gamma \int
\frac{d^3k}{(2\pi)^3}\, \hat{F}(\vec{k},t) \, \left[ e^{i \vec{k}
\cdot \vec{q}}\, \rho \, e^{-i \vec{k} \cdot \vec{q}} \; - \; \rho
\right].
\end{equation}

The above expression falls into the general class of
translational-invariant Markovian master equations first given by
Holevo~\cite{1993-h} which, in the case of a bounded mapping
${\mathcal L}$, reads:
\begin{equation} \label{eq:hol}
{\mathcal L} [\rho] = \int d \mu(\vec{k}) \sum_{n=1}^{\infty} \left[
e^{i \vec{k} \cdot \vec{q}}\, L_{n}(\vec{k},\vec{p}\,)\, \rho \,
L_{n}^{\dagger}(\vec{k},\vec{p}\,)\, e^{-i \vec{k} \cdot \vec{q}} \; -
\, \frac{1}{2} \left\{L_{n}^{\dagger}(\vec{k},\vec{p}\,)
L_{n}(\vec{k},\vec{p}\,), \rho \right\} \right],
\end{equation}
where $L_{n}(\vec{k},\vec{p}\,)$ are bounded functions of the momentum
operator $\vec{p}$, and $\mu(\vec{k})$ is a positive $\sigma$-finite
measure. Briefly, the physical content of \eqref{eq:hol} is the
following: the unitary operators $e^{i \vec{k} \cdot \vec{q}}$ and
$e^{-i \vec{k} \cdot \vec{q}}$ induce a momentum transfer to the
particle by an amount equal to $\vec{k}$, while the operators
$L_{n}(\vec{k},\vec{p}\,)$ imply that the momentum transfer to the
particle depends on the momentum of the particle itself. This allows
for mechanisms such as relaxation to take place.

Equation \eqref{eq:hol} reduces to~\eqref{eq:me} under the following
circumstances. Let us assume that $L_{n}(\vec{k},\vec{p}\,) =
L_{n}(\vec{k})$ does not depend on the momentum $\vec{p}$ of the
particle. They then become $c$-number functions, commuting with all
other operators. By setting:
\begin{equation}
d \mu (\vec{k}) \sum_{n=1}^{\infty} | L_{n}(\vec{k}) |^2 \; = \; 2
m^2 \gamma\, \frac{d^3k}{(2\pi)^3}\, \hat{F}(\vec{k},t),
\end{equation}
the link is established. Of course, in the truly Markovian case one
has $D(\vec{x} - \vec{y}, t-s) = G(\vec{x}~-~\vec{y}) \delta(t~-~s)$
so that $\hat{F}(\vec{k},t) = (1/2) \hat{G}(\vec{k})$ is independent
of time, where $\hat{G}(\vec{k})$ is the Fourier transform of
$G(\vec{x}~-~\vec{y})$.

According to the above analysis, the effect of the thermal field is
not only that of localizing the wave function in space (this is a
consequence of the specific form of the stochastic
equation~\eqref{eq:cm5}), but also of exchanging momentum between
the particle and the field. This is the reason why both the momentum
and the energy of the particle are not conserved, in general. One
would expect the energy of the particle to thermalize to that of the
random field; however, the model described by \eqref{eq:cm5}
does not allow for thermalization, since the operators
$L_{n}(\vec{k})$ do not depend on the momentum $\vec{p}$ of the
particle. This is in agreement with the results of Sec. 5A on
energy production. The comparison with decoherence suggests how the
model can be modified in order to include also such an effect; this
will be a subject of future research.

\section{Integrals in the dilute, nonrelativistic thermal model}

{}From \eqref{eq:intform} we find
\begin{align}\label{eq:ints1}
&\int_{-\infty}^{\infty}\frac{k^2dk}{(2\pi)^2}
  e^{- k^2/(2\mu T)}\frac{\sin(kR)}{kR}
  =\Big(\frac{\mu T}{2\pi}\Big)^{3/2}
  e^{-(\mu T R^2/2)}~~~,\cr
&\int_{-\infty}^{\infty}\frac{k^2dk}{(2\pi)^2}
  e^{- k^2/(2\mu T)}\frac{\sin(kR)}{kR} \exp\Big(i\big(\mu+\frac {
  k^2}{2\mu})t\big)\Big)\cr
  =&\Big(\frac{\mu T}{2\pi}\Big)^{3/2} (1+t^2T^2)^{-3/4}
  e^{-(\mu T R^2/2)/(1+t^2T^2)}\cr
  &\times \exp\Big(i\big(\mu t+(3/2)\tan^{-1}(tT)-(\mu t T^2R^2/2)/(1+t^2T^2)\big)\Big)~~~,\cr
\end{align}
from which, by forming linear combinations, taking real and
imaginary parts, and taking limits as $R=|\vec x|\to 0$, we get
\begin{align}\label{eq:ints2}
D(\vec 0,t)- D(\vec x,t)\simeq& \frac{e^{-(\mu-\zeta)/T}}{\mu}
\Big(\frac{\mu T}{2\pi}\Big)^{3/2} (1+t^2T^2)^{-3/4}
 \Big[\cos\Big(\mu t+(3/2)\tan^{-1}(tT)\Big)\cr
 - &e^{-(\mu T R^2/2)/(1+t^2T^2)} \cos\Big(\mu t+(3/2)\tan^{-1}(tT)-(\mu t T^2R^2/2)/(1+t^2T^2)\Big)\Big]~~~,\cr
F(\vec 0,t)-F(\vec x,t)\simeq& \frac{ e^{-(\mu-\zeta)/T}}{\mu^2}
\Big(\frac{\mu T}{2\pi}\Big)^{3/2} (1+t^2T^2)^{-3/4}
 \Big[\sin\Big(\mu t+(3/2)\tan^{-1}(tT)\Big)\cr
 -& e^{-(\mu T R^2/2)/(1+t^2T^2)}\sin\Big(\mu t+(3/2)\tan^{-1}(tT)-(\mu t
T^2R^2/2)/(1+t^2T^2)\Big)\Big]~~~,\cr
 I(\vec 0,t)-I(\vec x,t)\simeq& \frac{
e^{-(\mu-\zeta)/T}}{\mu^3}\Big(\frac{\mu
T}{2\pi}\Big)^{3/2}\Big\{1-e^{-(\mu T R^2/2)} \cr -&
 (1+t^2T^2)^{-3/4}
 \Big[\cos\Big(\mu t+(3/2)\tan^{-1}(tT)\Big)\cr
 -& e^{-(\mu T R^2/2)/(1+t^2T^2)}\cos\Big(\mu t+(3/2)\tan^{-1}(tT)-(\mu t
T^2R^2/2)/(1+t^2T^2)\Big)\Big]\Big\}~~~.\cr
\end{align}

\section{Time evolution of the wave function and exponential decay of
superpositions}

As mentioned in the introduction, an alternative form of the
collapse equation has been given in (I), which differs
from~\eqref{eq:cm5} by a change of measure for the noise; see
(35) and (37) of (I). The advantage of this alternative formulation
is that it can be expressed in terms of a {\it linear}, but {\it
not} norm-preserving, equation ( (34) of (I)), which is simpler
to solve. Upon normalization and change of measure, one recovers the
usual collapse dynamics.

Let us specialize to the case of a single particle; let us
moreover set $H = 0$, as we want to focus only on the collapse
mechanics. Then, for the mass density coupling considered in this
paper, the linear equation reads:
\begin{equation}\label{eq:xccg}
\frac{d|\chi(t)\rangle}{dt} = \left[ \sqrt{\gamma} \int d^3 x \,
M(\vec{x}) \phi(\vec{x},t) - 2 \gamma \int d^3 x \int d^3 y
M(\vec{x}) M(\vec{y}) F(\vec{x}- \vec{y},t) \right] |\chi(t)
\rangle~~~.
\end{equation}
The random field $\phi(\vec{x},t)$ is now supposed to be a Gaussian
thermal field with respect to a new measure ${\mathbb Q}$, having
mean 0 and correlator $D(\vec{x} - \vec{y}, t - s)$. The relation
between the statistical averages with respect to this measure and
the averages with respect to the physical measure used throughout this paper
(which we shall
call ${\mathbb P}$ from now on) is:
\begin{equation} \label{eq:fd}
{\mathbb E}_{\mathbb P} [ f(t) ] \; = \; {\mathbb E}_{\mathbb Q} [
f(t) \langle \chi(t) | \chi(t) \rangle ]~~~,
\end{equation}
where $f(t)$ is a generic random function of time.

Because of the special form~\eqref{eq:massden} of the particle
density operator $M(\vec{x})$, \eqref{eq:xccg} can be readily
solved in the coordinate representation $\chi(\vec{x}, t) = \langle
\vec{x} | \chi(t) \rangle$:
\begin{equation} \label{eq:xxc}
\chi(\vec{x}, t) = \exp\left[\sqrt{\gamma} m \Phi(\vec{x},t) - 2
\gamma m^2 I(\vec{0},t)\right]\, \chi(\vec{x}, 0)~~~,
\end{equation}
with:
\begin{equation}
\Phi(\vec{x},t) = \int_{0}^{t} ds \, \phi(\vec{x},s), \qquad
I(\vec{0},t) = \int_{0}^{t} ds \, F(\vec{0},s)
\end{equation}
($I(\vec{x},t)$ has been first introduced in \eqref{eq:redint}.)
Let us fix an arbitrary time $t$. Then the random field
$\Phi(\vec{x},t)$ is a Gaussian field in the variable $\vec{x}$,
with mean and correlator equal to:
\begin{equation} \label{eq:xzxzv}
{\mathbb E}_{\mathbb Q} [ \Phi(\vec{x},t) ] = 0, \qquad {\mathbb
E}_{\mathbb Q} [ \Phi(\vec{x},t) \Phi(\vec{y},t) ] = 2 I(\vec{x} -
\vec{y},t)~~~.
\end{equation}
The above statistical properties refer to the measure ${\mathbb Q}$,
while we need them to be expressed with respect to the physical
measure ${\mathbb P}$. \eqref{eq:fd} allows us to switch between
the two measures, once the squared norm $\langle \chi(t) | \chi(t)
\rangle$ has been computed.

In analogy with the discussion of Sec. 3, let us consider an
initial state of the form:
\begin{equation} \label{eq:zxp}
\chi(\vec{x}, 0) = \alpha_{1} \delta^3(\vec{x} - \vec{r}^{\,1})^{1/2}
+ \alpha_{2} \delta^3(\vec{x} - \vec{r}^{\,2})^{1/2}~~~,
\end{equation}
corresponding to the superposition of two states well localized
around $\vec{r}^{\,1}$ and $\vec{r}^{\,2}$ respectively. By substituting
it into \eqref{eq:xxc} and normalizing the wave function, one
obtains for the collapse probabilities:
\begin{eqnarray}
p_{1}(t) = |\alpha_{1}(t)|^2 & = & \frac{p_{1} e^{2\sqrt{\gamma} m
\Phi(\vec{r}^{\,1},t)}}{p_{1} e^{2\sqrt{\gamma} m \Phi(\vec{r}^{\,1},t)} +
p_{2} e^{2\sqrt{\gamma} m \Phi(\vec{r}^{\,2},t)}}~~~, \nonumber\\
p_{2}(t) = |\alpha_{1}(t)|^2 & = & \frac{p_{2} e^{2\sqrt{\gamma} m
\Phi(\vec{r}^{\,2},t)}}{p_{1} e^{2\sqrt{\gamma} m \Phi(\vec{r}^{\,1},t)} +
p_{2} e^{2\sqrt{\gamma} m \Phi(\vec{r}^{\,2},t)}}~~~, \label{eq:zssx}
\end{eqnarray}
with $p_{1} = |\alpha_{1}|^2$ and $p_{2} = |\alpha_{2}|^2$. Using
\eqref{eq:fd}, together with the equation
\begin{eqnarray} \label{eq:xfktr}
\langle \chi(t) | \chi(t) \rangle & = & p_{1} \exp \left[ 2
\sqrt{\gamma} m \Phi(\vec{r}^{\,1},t) - 4 \gamma m^2 I(\vec{0},t)
\right] \nonumber \\
& + & p_{2} \exp \left[ 2 \sqrt{\gamma} m \Phi(\vec{r}^{\,2},t) - 4
\gamma m^2 I(\vec{0},t) \right]~~~,
\end{eqnarray}
we can compute the average of the product $p_{1}(t)p_{2}(t)$.

Due to the statistical properties~\eqref{eq:xzxzv}, the joint
probability density of the two random variables
$\Phi(\vec{r}^{\,1},t)$ and $\Phi(\vec{r}^{\,2},t)$ reads:
\begin{equation} \label{eq:zhf}
P^{\mathbb Q}_{12} = \frac{1}{2 \pi \sqrt{a_t^2 - b_t^2}} \exp
\left[ - \frac{a_t(\Phi(\vec{r}^{\,1},t))^2 - 2a_t
b_t\Phi(\vec{r}^{\,1},t)\Phi(\vec{r}^{\,2},t) +
a_t(\Phi(\vec{r}^{\,2},t))^2}{2(a_t^2 - b_t^2)} \right] ~~~,
\end{equation}
with $a_t = 2 I(\vec{0},t)$ and $b_t = 2 I(\vec{r}^{\,1} -
\vec{r}^{\,2},t)$. Using now
\eqref{eq:fd},~\eqref{eq:zssx},~\eqref{eq:xfktr} and~\eqref{eq:zhf}
we get:
\begin{equation} \label{eq:fdfp}
{\mathbb E}_{\mathbb P} [p_{1}(t)p_{2}(t)] = p_{1}p_{2} e^{-2 \gamma
m^2 a_t} \, \frac{1}{8 \pi \gamma m^2 a_t \sqrt{1 - r_t^2}} \,
\int_{-\infty}^{+\infty} dx \, dy \frac{\exp\left[-\frac{x^2 + y^2
-2 r_t xy}{8\gamma m^2 a_t(1 - r_t^2)} +x+y\right]}{p_{1} e^{x} +
p_{2} e^{y}}~~~,
\end{equation}
with $r_t = b_t/a_t$; we have also relabeled $x = 2 \sqrt{\gamma} m
\Phi(\vec{r}^{\,1},t)$ and $y = 2 \sqrt{\gamma} m
\Phi(\vec{r}^{\,2},t)$. \eqref{eq:fdfp} can be further simplified by
making the change of variables $t = (x+y)/2$, $s = (x-y)/2$. In this
case, the two integrals decouple and one gets:
\begin{equation} \label{eq:fdfp2}
{\mathbb E}_{\mathbb P} [p_{1}(t)p_{2}(t)] = p_{1}p_{2} e^{-
\Gamma(t)} \, \frac{1}{2 \sqrt{\pi \Gamma(t)}} \,
\int_{-\infty}^{+\infty} ds \frac{e^{-s^2/4\Gamma(t)}}{p_{1} e^{s} +
p_{2} e^{-s}}~~~,
\end{equation}
where $\Gamma(t) = \gamma m^2 a_t (1-r_t)$ corresponds to the
definition~\eqref{eq:gamnew}. The final integral gives a finite
contribution as $\Gamma(t) \to \infty$, which proves that the decay
of the superposition is exponential in time, and proportional to
$e^{- \Gamma(t)}/\sqrt{\Gamma(t)}$. In particular, by using the
inequality
\begin{equation}
p_{1} e^{s} + p_{2} e^{-s} \; \geq \; \overline{m} (e^{s} + e^{-s})
\; = \; 2 \overline{m} \cosh s~~~,
\end{equation}
with $\overline{m} \equiv \min \{ p_{1}, p_{2} \}$ (here we assume
that $\overline{m} \neq 0$; the trivial case $\overline{m}=0$  can be treated
separately), one has:
\begin{equation}
\int_{-\infty}^{+\infty} ds \frac{e^{-s^2/4\Gamma(t)}}{p_{1} e^{s} +
p_{2} e^{-s}} \; \leq \; \frac{1}{2 \overline{m}}
\int_{-\infty}^{+\infty} ds \frac{1}{\cosh s} \; = \; \frac{\pi}{2
\overline{m}}~~~;
\end{equation}
Collecting all results, we can write:
\begin{equation}
{\mathbb E}_{\mathbb P} [p_{1}(t)p_{2}(t)] \; \leq \; {\mathbb
E}_{\mathbb P}
[p_{1}(0)p_{2}(0)]\,\frac{\sqrt{\pi}}{4\overline{m}\sqrt{\Gamma(t)}}\,
e^{- \Gamma(t)}~~~.
\end{equation}

\section{Unparticle thermal correlation functions}

We take the unparticle thermal correlation function to be given by an average
over thermal correlation functions for particles of mass $\mu\geq 0$, using the same
weighting function $\rho(\mu^2)$ that is used \cite{kras} to generate the unparticle propagator
from the propagator for a boson of mass $\mu$,
\begin{equation}\label{eq:weight}
\rho(\mu^2)=(d-1) \Lambda^{2(1-d)}(\mu^2)^{d-2}~~~.
\end{equation}
Writing the left hand side of \eqref{eq:dsimp} as $D(\vec x,t,\mu)$ so
as to explicitly show the mass dependence, the thermal unparticle correlation function $D_{\cal U}$ is then given by
\begin{equation}\label{eq:undef}
D_{\cal U}(\vec x,t)= \int_0^{\infty} d\mu^2 \rho(\mu^2)
D(\vec x,t,\mu)~~~.
\end{equation}
Substituting \eqref{eq:dsimp} and \eqref{eq:ndef1}, we thus get
\begin{align}\label{eq:unintegral}
D_{\cal U}(\vec x,t)=&
(d-1) \Lambda^{2(1-d)} \int_0^{\infty} d\mu^2 (\mu^2)^{d-2}
\int\frac{ d^3 k}{(2\pi)^3 \omega_k}\frac{1} {e^{\frac {\omega_k-\zeta }{T}}-1 }
 \cos\big(\vec k \cdot \vec x \big)\cos\big(\omega_k t\big)\cr
=&(d-1) \Lambda^{2(1-d)} \int\frac{ d^3 k}{(2\pi)^3}\cos\big(\vec k \cdot \vec x \big)\int_0^{\infty} d\mu^2 (\mu^2)^{d-2} \frac{1}{\omega_k}\frac{1} {e^{\frac {\omega_k-\zeta }{T}}-1 }\cos\big(\omega_k t\big)~~~,\cr
\end{align}
where in the second line we have isolated those factors of the integrand that
explicitly depend on $\mu$.  Since $\omega_k^2=k^2+\mu^2$, we can change integration
variable in the inner integral from $\mu^2$ to $\omega_k^2$, by using
\begin{equation}\label{eq:changvar}
d\mu^2=2\omega_k d\omega_k~~~,~~(\mu^2)^{d-2}=(\omega_k^2-k^2)^{d-2}~~~,
\end{equation}
which gives
\begin{equation}\label{eq:inner1}
\int_0^{\infty} d\mu^2 (\mu^2)^{d-2} \frac{1}{\omega_k}\frac{1} {e^{\frac {\omega_k-\zeta }{T}}-1 }\cos\big(\omega_k t\big)
=2\int_k^{\infty} d\omega\frac{(\omega^2-k^2)^{d-2}} {e^{\frac {\omega-\zeta }{T}}-1 }\cos\big(\omega t\big)~~~,
\end{equation}
where we have relabeled the dummy integration variable $\omega_k$ as $\omega$.
Substituting this into \eqref{eq:unintegral} we get
\begin{equation}\label{eq:unintegral1}
D_{\cal U}(\vec x,t)=2(d-1) \Lambda^{2(1-d)} \int\frac{ d^3 k}{(2\pi)^3}\cos\big(\vec k \cdot \vec x \big) \int_k^{\infty} d\omega\frac{(\omega^2-k^2)^{d-2}} {e^{\frac {\omega-\zeta }{T}}-1 }\cos\big(\omega t\big)~~~.
\end{equation}
Corresponding to this formula, the function $F_{\cal U}(\vec x,t)$
introduced in \eqref{eq:fdeff} is given by
\begin{equation}\label{eq:fintegral1}
F_{\cal U}(\vec x,t)=\int_0^tds D_{\cal U}(\vec x,t-s) =
2(d-1) \Lambda^{2(1-d)} \int\frac{ d^3 k}{(2\pi)^3}\cos\big(\vec k \cdot \vec x \big) \int_k^{\infty}\frac{ d\omega}{\omega}\frac{(\omega^2-k^2)^{d-2}} {e^{\frac {\omega-\zeta }{T}}-1 }\sin\big(\omega t\big)~~~,
\end{equation}
and the integral appearing in the  rate function $\Gamma(t)$ of
\eqref{eq:gamdef} is given by
\begin{equation}\label{eq:redint1}
I_{\cal U}(\vec x,t)\equiv \int_0^t ds F_{\cal U}(\vec x,s)
=2(d-1) \Lambda^{2(1-d)} \int\frac{ d^3 k}{(2\pi)^3}\cos\big(\vec k \cdot \vec x \big) \int_k^{\infty}\frac{ d\omega}{\omega^2}\frac{(\omega^2-k^2)^{d-2}} {e^{\frac {\omega-\zeta }{T}}-1 }\big[1-\cos\big(\omega t\big)\big]~~~.
\end{equation}
{}From \eqref{eq:fintegral1} we can read off the Fourier transform defined in
\eqref{eq:fourrep}, from which the energy production is calculated through
\eqref{eq:enprod4},
\begin{equation}\label{eq:four1}
\hat F_{\cal U}(\vec k,t) =
 2(d-1) \Lambda^{2(1-d)}  \int_k^{\infty}\frac{ d\omega}{\omega}\frac{(\omega^2-k^2)^{d-2}} {e^{\frac {\omega-\zeta }{T}}-1 }\sin\big(\omega t\big)~~~.
\end{equation}
Note that in all of these formulas, the scale parameter $\Lambda$ appears as an overall factor,
which then combines with the noise coupling $\gamma$ to give a new effective coupling $\gamma \Lambda^{2(1-d)}$.

The correlation function $D_{\cal U}(\vec x,t)$ can be written in several alternative forms.  Performing the angular average over $\vec k$, we get
\begin{equation}\label{eq:unintegral2}
D_{\cal U}(\vec x,t)=(d-1) \Lambda^{2(1-d)} \int_0^{\infty} \frac{ dk \, k} {\pi^2|\vec x|}\sin\big(k |\vec x| \big) \int_k^{\infty} d\omega\frac{(\omega^2-k^2)^{d-2}} {e^{\frac {\omega-\zeta }{T}}-1 }\cos\big(\omega t\big)~~,
\end{equation}
which on interchange of orders of the $k$ and $\omega$ integrations becomes
\begin{equation}\label{eq:unintegral3}
D_{\cal U}(\vec x,t)=(d-1) \Lambda^{2(1-d)}
\int_0^{\infty} d\omega \frac{\cos\big(\omega t\big)} {e^{\frac {\omega-\zeta }{T}}-1 }
\int_0^{\omega} \frac{ dk \, k} {\pi^2|\vec x|}\sin\big(k |\vec x| \big) (\omega^2-k^2)^{d-2}~~~.
\end{equation}
Making the change of integration variable $k=\omega v$, this can be further
rewritten as
\begin{equation}\label{eq:unintegral4}
D_{\cal U}(\vec x,t)=(d-1) \Lambda^{2(1-d)}
\int_0^{\infty} d\omega \frac{\omega^{2(d-1)}\cos\big(\omega t\big)} {e^{\frac {\omega-\zeta }{T}}-1 }
\int_0^1 \frac{ dv \, v} {\pi^2|\vec x|}\sin\big(v \omega |\vec x| \big) (1-v^2)^{d-2}~~~.
\end{equation}
The integral over $v$ in \eqref{eq:unintegral4} converges only for ${\rm Re}\,d>1$.  However,
by an integration by parts this integral is transformed as follows,
\begin{equation}\label{eq:intparts}
\int_0^1 \frac{ dv \, v} {\pi^2|\vec x|}\sin\big(v \omega |\vec x| \big) (1-v^2)^{d-2}=\int_0^1 \frac{dv \omega }{2 \pi^2 (d-1)}\cos\big(v \omega |\vec x|\big) (1-v^2)^{d-1}~~~,
\end{equation}
which gives an analytic continuation around the simple pole at $d=1$, expressed in terms of  a
$v$ integral that now converges for ${\rm Re}\,d>0$.  Substituting \eqref{eq:intparts} into
\eqref{eq:unintegral4} gives a formula for the correlation function which is now
manifestly finite for ${\rm Re}\,d>0$,
\begin{equation}\label{eq:unintegral5}
D_{\cal U}(\vec x,t)=\frac{1}{2} \Lambda^{2(1-d)}
\int_0^{\infty} d\omega \frac{\omega^{2d-1}\cos\big(\omega t\big)} {e^{\frac {\omega-\zeta }{T}}-1 }
\int_0^1 \frac{ dv} {\pi^2}\cos\big(v \omega |\vec x| \big) (1-v^2)^{d-1}~~~.
\end{equation}
The corresponding formulas for  $F_{\cal U}(\vec x,t)$ and $I_{\cal U}(\vec x,t)$ are now obtained by  the
replacement of $\cos(\omega t)$ by $\sin(\omega t)/\omega $ and $[1-\cos(\omega t)]/
\omega^2$, respectively,
\begin{equation}\label{eq:fint1}
F_{\cal U}(\vec x,t)=\frac{1}{2} \Lambda^{2(1-d)}
\int_0^{\infty} d\omega \frac{\omega^{2d-2}\sin\big(\omega t\big)} {e^{\frac {\omega-\zeta }{T}}-1 }
\int_0^1 \frac{ dv} {\pi^2}\cos\big(v \omega |\vec x| \big) (1-v^2)^{d-1}~~~,
\end{equation}
and
\begin{equation}\label{eq:iint1}
I_{\cal U}(\vec x,t)=\frac{1}{2} \Lambda^{2(1-d)}
\int_0^{\infty} d\omega \frac{\omega^{2d-3}\big[1-\cos\big(\omega t\big)\big]} {e^{\frac {\omega-\zeta }{T}}-1 }
\int_0^1 \frac{ dv} {\pi^2}\cos\big(v \omega |\vec x| \big) (1-v^2)^{d-1}~~~.
\end{equation}
{}From \eqref{eq:iint1} we find for the subtracted integral that enters into
$\Gamma(t)$,
\begin{equation}\label{eq:iint2}
I_{\cal U}(\vec 0,t)-I_{\cal U}(\vec x,t)=\frac{1}{2} \Lambda^{2(1-d)}
\int_0^{\infty} d\omega \frac{\omega^{2d-3}\big[1-\cos\big(\omega t\big)\big]} {e^{\frac {\omega-\zeta }{T}}-1 }
\int_0^1 \frac{ dv} {\pi^2}\big[1-\cos\big(v \omega |\vec x| \big)\big] (1-v^2)^{d-1}~~~,
\end{equation}
giving an expression that is manifestly positive.

Let us now return to  \eqref{eq:four1}, and use it to calculate the energy
production.  Substituting \eqref{eq:four1} into \eqref{eq:enprod4} we get, for a single particle with mass-coupled unparticle noise,
\begin{equation}\label{eq:enprod5b}
\frac{d}{dt}{\rm Tr}H\rho(t)=\frac{\gamma m \Lambda^{2(1-d)}(d-1)}{\pi^2}
\int_0^{\infty} dk\, k^4 \int_k^{\infty}
\frac{ d\omega}{\omega}\frac{(\omega^2-k^2)^{d-2}} {e^{\frac {\omega-\zeta }{T}}-1 }\sin\big(\omega t\big)~~~,
\end{equation}
which on interchanging the orders of the $k$ and $\omega$ integrations becomes
\begin{equation}\label{eq:enprod6b}
\frac{d}{dt}{\rm Tr}H\rho(t)
=\frac{\gamma m \Lambda^{2(1-d)}(d-1)}{\pi^2}
\int_0^{\infty}
\frac{ d\omega}{\omega}\frac{\sin\big(\omega t\big)} {e^{\frac {\omega-\zeta }{T}}-1 }\int_0^{\omega} dk\, k^4(\omega^2-k^2)^{d-2} ~~~.
\end{equation}
Making the change of variable
$k=\omega u^{1/2}$, $dk=(1/2)\omega u^{-1/2}du$ in the inner integral, it
can be evaluated in terms of the Euler $B$ function; then using
$(d-1)\Gamma(d-1)=\Gamma(d)$  one gets the compact expression
\begin{equation}\label{eq:enratefinal}
\frac{d}{dt}{\rm Tr}H\rho(t)
=\frac{3\gamma m\Lambda^{2(1-d)}}{(2\pi)^2}\frac{\Gamma(3/2)\Gamma(d)}
{\Gamma(3/2+d)}
\int_0^{\infty}d\omega \frac {\omega^{2d} \sin\big(\omega t\big)}
{e^{\frac {\omega-\zeta }{T}}-1 }~~~,
\end{equation}
with the integral convergent for ${\rm Re}\,d>0$.
Integrating over t, one gets the corresponding
formula for the total energy production,
\begin{equation}\label{eq:enprodfinal}
{\rm Tr}H\rho(t)-{\rm Tr}H\rho(0)
=\frac{3\gamma m\Lambda^{2(1-d)}}{(2\pi)^2}\frac{\Gamma(3/2)\Gamma(d)}
{\Gamma(3/2+d)}
\int_0^{\infty}d\omega \frac {\omega^{2d-1}\big[1- \cos\big(\omega t\big)]}
{e^{\frac {\omega-\zeta }{T}}-1 }~~~.
\end{equation}

\end{document}